\documentclass[journal,twocolumn,10pt]{IEEEtran}
\usepackage{amsmath,amsfonts}
\usepackage{algorithmic}
\usepackage{array}
\usepackage{textcomp}
\usepackage{stfloats}
\usepackage{url}
\usepackage{verbatim}
\usepackage{graphicx}
\usepackage{balance}

\usepackage{subfigure}
\usepackage{multirow}
\usepackage{algorithm}
\usepackage{siunitx}
\usepackage{enumitem}

\usepackage{soulutf8}
\RequirePackage[normalem]{ulem} 
\RequirePackage{color}\definecolor{RED}{rgb}{1,0,0}\definecolor{BLUE}{rgb}{0,0,1} 

\def\BibTeX{{\rm B\kern-.05em{\sc i\kern-.025em b}\kern-.08em
    T\kern-.1667em\lower.7ex\hbox{E}\kern-.125emX}}

\begin{document}
\title{Boosted Neural Decoders: Achieving Extreme Reliability of LDPC Codes for 6G Networks}

\author{Hee-Youl Kwak, \IEEEmembership{Member,~IEEE}, 
Dae-Young Yun,~Yongjune Kim, \IEEEmembership{Member,~IEEE}, Sang-Hyo Kim, \IEEEmembership{Member,~IEEE}, and Jong-Seon~No, \IEEEmembership{Fellow,~IEEE}

\thanks{An earlier version of this paper was presented in part at the Annual Conference on Neural Information Processing Systems, New Orleans, LA, USA, Dec. 2023~\cite{Kwak2023}. (Corresponding authors: Y.~Kim and S.-H.~Kim.)}
\thanks{Hee-Youl Kwak is with the Department of Electrical, Electronic and
Computer Engineering, University of Ulsan, Ulsan 44610, South Korea. (e-mail: ghy1228@gmail.com).}
\thanks{Dae-Young~Yun is with Samsung Electronics, Co., Ltd., Hwasung 18448, South Korea (e-mail: bigbowl204@snu.ac.kr).}
\thanks{Yongjune Kim is with the Department of Electrical Engineering, Pohang University of Science and Technology, Pohang 37673, South Korea, and also with the Institute for Convergence Research and Education in
Advanced Technology, Yonsei University, Seoul 03722, South Korea (e-mail: yongjune@postech.ac.kr).}
\thanks{Sang-Hyo~Kim is with the Department of Electrical and Computer Engineering, Sungkyunkwan University, Suwon 16419, South Korea (e-mail: iamshkim@skku.edu).}
\thanks{Jong-Seon~No is with the Department of Electrical and Computer Engineering, INMC, Seoul National University, Seoul 08826, South Korea (e-mail: jsno@snu.ac.kr).}

}


\maketitle

\begin{abstract}
Ensuring extremely high reliability in channel coding is essential for 6G networks. The next-generation of ultra-reliable and low-latency communications (xURLLC) scenario within 6G networks requires frame error rate (FER) below $10^{-9}$. However, low-density parity-check (LDPC) codes, the standard in 5G new radio (NR), encounter a challenge known as the error floor phenomenon, which hinders to achieve such low rates. To tackle this problem, we introduce an innovative solution: boosted neural min-sum (NMS) decoder. This decoder operates identically to conventional NMS decoders, but is trained by novel training methods including: i) boosting learning with uncorrected vectors, ii) block-wise training schedule to address the vanishing gradient issue, iii) dynamic weight sharing to minimize the number of trainable weights, iv) transfer learning to reduce the required sample count, and v) data augmentation to expedite the sampling process. Leveraging these training strategies, the boosted NMS decoder achieves the state-of-the art performance in reducing the error floor as well as superior waterfall performance. Remarkably, we fulfill the 6G xURLLC requirement for 5G LDPC codes without a severe error floor. Additionally, the boosted NMS decoder, once its weights are trained, can perform decoding without additional modules, making it highly practical for immediate application. The source code is available at \mbox{\url{https://github.com/ghy1228/LDPC_Error_Floor}.}
\end{abstract}


\section{Introduction}
\IEEEPARstart{A}{chieving} extremely high reliability, characterized by low frame error rate (FER) of $10^{-9}$, is pivotal for realizing applications in the next generation of ultra-reliable and low-latency communications (xURLLC) within 6G networks \cite{Tataria2021,hong2022}. 
In particular, this level of reliability is crucial for mission-critical applications such as industrial automation, telemedicine, and autonomous driving, serving as key performance indicators (KPIs) \cite{Yang2021}. 
From the perspective of channel coding, it is imperative to develop codes and decoders that achieve extreme reliability with low signal to noise ratios (SNRs). 
This high reliability is also required for various systems such as solid-state drive (SSD) storage \cite{dong2010soft}, DNA storage \cite{chandak2019improved}, and cryptosystems \cite{baldi2008new}.


However, the low-density parity-check (LDPC) codes employed in the data channel of 5G new radio (NR) suffer from the error floor phenomenon, which hinders the achievement of extremely low FER at low SNRs \cite{Richardson2003}. 
The FER performance curve of LDPC codes typically comprises two distinct regions: the waterfall region, characterized by a sharp decline in FER, and the error floor region, where the decrease in FER becomes more gradual. 
Techniques to improve waterfall performance often inadvertently worsen error floor performance. For example, introducing irregularity into the code structure or employing decoders with weighting (e.g., weighted min-sum: WMS) can improve waterfall performance but degrade error floor performance. 
Moreover, the implementation of quantization, which is essential for efficient hardware realization, can lead to a severe error floor \cite{zhang2014quantized}.

\subsection{Main Contributions}
To mitigate the error floor, this paper proposes novel {\em training} methods for the neural min-sum (NMS) decoder \cite{nachmani2018deep}. As a generalized version of the WMS decoder, the NMS decoder incorporates multiple trainable weights that are trained using machine learning algorithms. We show that the error floor can be significantly reduced by the following five training methods.

 \begin{enumerate}[leftmargin=*]
\item {\em Boosting learning using uncorrected vectors:} 
We first leverage the boosting learning technique \cite{freund1997decision} from the machine learning domain. This technique involves a sequential training approach for multiple classifiers, where each classifier concentrates on samples incorrectly classified by its preceding classifiers. 
Following this approach, we split the whole decoding process into two stages: base and post. The post stage decoder is trained on specific received vectors that remain uncorrected in the base stage. If these uncorrected (UC) vectors are collected in the error floor region of the base stage, the post stage decoder is specifically trained to correct these UC vectors causing the error floor.

\item{\em Block-wise training schedule:} 
Reducing the error floor typically necessitates a large maximum number of iterations \cite{zhang2014quantized, Hatami2020, han2022deep}, potentially leading to the vanishing gradient problem \cite{glorot2010understanding} as a common challenge in deep networks. To overcome this issue, we propose a block-wise training schedule that sequentially trains sub-iteration blocks. We also retrain the weights from previous blocks in a sliding window manner to prevent getting stuck in local minima.

\item{\em Dynamic weight sharing:} 
We introduce a dynamic weight sharing technique that assigns distinct weights to unsatisfied check nodes (UCNs) and satisfied check nodes (SCNs) dynamically during the decoding process, and shares weights within the same iteration. This method significantly reduces the number of trainable weights while preserving performance due to the increased weight diversity from dynamic allocation.

\item{\em Transfer learning:}
Boosting learning, unlike conventional methods, requires numerous decoding trials to acquire UC vectors. 
To mitigate this time-consuming sampling process, transfer learning is employed, where initial weights are set by trained weights on different codes. Starting with these pre-trained weights, optimal weights can be obtained by fine-tuning with few samples. 
Transfer learning speeds up the sampling process by tens of times.

\item{\em Data augmentation:} We propose a data augmentation method based on importance sampling to further expedite the sampling process. Similar to the importance sampling technique used for predicting error floor performance \cite{Richardson2003, Dolecek2009}, we selectively amplify noise at error bit positions of given UC vectors. This method increases the probability of encountering UC vectors. By adopting the transfer learning and data augmentation methods, we are able to accelerate the sampling speed significantly.
\end{enumerate}

\begin{table*}[t]
\centering
    \caption{Comparison between model-based neural decoders}
    \label{Table:Compare_Conv}
\begin{tabular}{ccccccccc}
\hline
\multirow{2}{*}{Ref.}                     & \multirow{2}{*}{Codes}        & \multirow{2}{*}{Target region} & \multirow{2}{*}{Decoders} & Training                          & Training                    & Weight                    & Transfer  & Data           \\
                                          &                               &                                &                           & Sample                           & Schedule                    & Sharing                   & Learning      & Augmentation       \\ \hline \hline
\multirow{2}{*}{Ours}                     & \multirow{2}{*}{Standard LDPC}                      & Waterfall,                     & \multirow{2}{*}{MS}       & \multirow{2}{*}{UC vectors}   & \multirow{2}{*}{Block-wise} & \multirow{2}{*}{Dynamic}              & \multirow{2}{*}{Yes} & \multirow{2}{*}{Yes} \\
                                          &                           & Error floor                    &                           &                                   &                             &                &                      \\ \hline
\cite{nachmani2018deep}                   & BCH                           & Waterfall                      & BP, MS                    & Received vectors                  & One-shot                    & Temporal                  & No    & No               \\ \hline
\cite{dai2021learning}                    & Standard LDPC                 & Waterfall                      & BP, MS                    & Received vectors                  & Iter-by-Iter                & Spatial                   & No    & No               \\ \hline
\cite{rosseel2022decoding}                & Short LDPC                    & Waterfall                      & BP                        & Absorbing set                     & One-shot                    & Temporal                  & No       & No            \\ \hline
\multirow{2}{*}{\cite{xiao2020designing}} & \multirow{2}{*}{Regular LDPC} & Waterfall,                     & \multirow{2}{*}{FAID}     & \multirow{2}{*}{Received vectors} & \multirow{2}{*}{One-shot}   & \multirow{2}{*}{Temporal} & \multirow{2}{*}{No}  & \multirow{2}{*}{No} \\
                                          &                               & Error floor                    &                           &                                   &                             &                           &                      \\ \hline
\multirow{2}{*}{\cite{xiao2021faid}}      & \multirow{2}{*}{Short LDPC}   & Waterfall,                     & \multirow{2}{*}{FAID}     & \multirow{2}{*}{Trapping set}     & \multirow{2}{*}{One-shot}   & \multirow{2}{*}{Temporal} & \multirow{2}{*}{No} & \multirow{2}{*}{No}  \\
                                          &                               & Error floor                    &                           &                                   &                             &                           &                      \\ \hline
\cite{shah2021neural}                     & Standard LDPC                 & Waterfall                      & Layered                   & Received vectors                  & Iter-by-Iter                & Spatial               & No    & No     \\
\hline
\end{tabular}
\end{table*}

We refer to the NMS decoder trained through these methods as the boosted NMS decoder.
For various practical LDPC codes used in WiMAX \cite{ieee80216e2020}, WiFi \cite{ieee802112016}, and 5G NR \cite{3gpp2023}, the boosted NMS decoder exhibits notable improvements in error floor performance compared to the WMS decoder \cite{chen2005reduced}, threshold attenuated MS (TAMS) decoder \cite{Hatami2020} and other NMS decoders \cite{nachmani2018deep, dai2021learning}. The boosted NMS decoder also attains comparable performance to the state-of-the-art post-processing method \cite{han2022deep}, while requiring only one-third the number of iterations. 
Importantly, this performance enhancement is achieved solely through modifications to the training process. Consequently, once the training is completed, the boosted NMS decoder can be seamlessly incorporated into the existing hardware architectures designed for the WMS decoder, without requiring additional hardware modifications.

This paper presents novel contributions building upon our previous work \cite{Kwak2023}, specifically:
\begin{enumerate}[leftmargin=*]
\item We propose the transfer learning method specialized for the boosted NMS decoder to reduce the training cost when training multiple codes.
\item We propose the data augmentation method aimed at accelerating the sampling process, which is essential for applying boosting learning at extremely low error rates.
\item While our previous work showed the achievement of FER around $10^{-7}$, this paper advances further by achieving the xURLLC FER target of $10^{-9}$ without severe error floors for 5G LDPC codes.
\item Further, this paper provides an analysis of the loss function and the flexibility, and includes comparisons with recent studies \cite{Hatami2020, rosseel2022decoding, choukroun2022error}, enhancing the understanding of the proposed decoder.
\end{enumerate}

\subsection{Related Works}
{\em Error Floor:} Error floors are typically associated with problematic small graph structures called trapping sets \cite{Richardson2003}. Therefore, prior research has focused on developing design algorithms to eliminate trapping sets \cite{Richter2006}--\cite{Karimi2021} and optimizing decoders to escape trapping sets through perturbations in the decoding process \cite{zhang2014quantized,Hatami2020}--\cite{lee2022post}.
The TAMS decoder \cite{Hatami2020} applies weighting based on the minimum message value for each check node and the post-processing decoder \cite{han2022deep} utilizes neural networks to obtain information for managing post-processing; however, they do not employ neural decoders.
The study \cite{zhang2014quantized} utilized a non-uniform quantization scheme, which is a complementary approach with ours as we will assume the uniform quantization.

{\em Neural Decoders:} Neural decoders refer to decoders that perform or aid the decoding task using neural networks. Initially, research focused on replacing decoders with fully connected networks \cite{caid1990neural}--\cite{Lee2020}, and subsequently, more advanced architectures, such as recurrent neural network (RNN) \mbox{\cite{Bennatan2018, Jiang2020}}, autoencoder \mbox{\cite{O’Shea2017}--\cite{Clausius2023}}, and transformer \mbox{\cite{choukroun2022error}}, were explored. Another significant research direction involves model-based neural decoders that employ structured neural networks derived from algorithms like belief propagation (BP) and message passing (MS) \mbox{\cite{nachmani2018deep}}.
Model-based neural decoders have primarily been applied to high-density codes for alleviating performance degradation caused by structural weaknesses such as cycles \mbox{\cite{nachmani2018deep,lian2019learned, Kumar2023}}. 
More recently, the application of model-based neural decoders has expanded to include a broader range of code classes such as LDPC \mbox{\cite{dai2021learning, rosseel2022decoding, Wang2024,Andreev2021}}, generalized LDPC \cite{Kwak2022}, cyclic codes \cite{Chen2021}, several decoder classes including finite alphabet iterative decoders (FAIDs) \cite{xiao2020designing, xiao2021faid}, layered decoders \cite{shah2021neural}, and soft-output decoders \mbox{\cite{Artemasov2023}}. Moreover, various attempts have been made to enhance the performance by leveraging machine learning techniques (e.g., hyper-graph \cite{nachmani2019hyper}, active learning \cite{Beery2020}, pruning \cite{Buchberger2021}, and knowledge distillation \cite{Nachmani2022}).

The comparison with various model-based neural decoders is shown in Table \ref{Table:Compare_Conv}. A notable point of comparison is the training sample. 
Most studies have used received vectors simply sampled from the additive white Gaussian noise (AWGN) channel. Recent studies \cite{rosseel2022decoding,xiao2021faid} generate noisy codeword vectors with intentionally introduced errors in trapping sets. However, this method requires identifying the trapping sets, which is highly complex and feasible only for short codes. In contrast, the proposed boosting method, which collects training samples through decoding of linear complexity, is applicable to longer codes. 
Moreover, the distribution of the training samples for the post decoder is identical to that of the received vectors encountered during the actual post decoding stage, which makes it a more intuitive and natural method.


In \cite{nachmani2019hyper}, the authors improved waterfall performance by integrating hypernetworks into the standard neural decoder architecture. A transformer architecture was investigated to decode several classes of short codes \cite{choukroun2022error}.
They achieved enhanced waterfall performance for short codes at the expense of increased training and decoding costs.
Although our proposed training techniques are broadly applicable to these augmented neural decoders, our work primarily focuses on improving error floor performance of moderate and long codes with the standard NMS decoder under practical conditions.
Instead, we provide a comparative analysis for short codes in the waterfall region, which demonstrates that the proposed methods are also effective for short LDPC codes.

The rest of the paper is organized as follows: Section II provides background preliminaries. Section III introduces our proposed training methods for enhancing the NMS decoder's resistance to the error floor. In Section IV, we evaluate the performance of the boosted NMS decoder. Concluding remarks are offered in Section V.

\section{Preliminaries}\label{Sec:Construction}

\subsection{LDPC Codes}
Quasi-cyclic (QC) LDPC codes are widely adopted across various applications due to their implementation benefits
\cite{ryan2009channel,fossorier2004quasicyclic}. 
We focus on QC-LDPC codes in this study to validate the practicality of our research. A QC-LDPC code is constructed by lifting a protograph, which consists of $M$ proto variable nodes (VNs), $N$ proto CNs, and $E$ proto edges, with a lifting factor~$z$ \cite{fossorier2004quasicyclic,thorpe2003low}. 
The resulting Tanner graph comprises $n=Nz$ VNs and $m=Mz$ CNs. As a running example, we use the WiMAX QC-LDPC code with the length $n=576$ and code rate $R=3/4$, which is lifted from a protograph with $N=24, M=6, E=88, z=24$ \cite{ieee80216e2020}.

\subsection{Neural Min-Sum Decoding}
Let ${\hat m}^{(\ell)}_{c\rightarrow v}~(m^{(\ell)}_{v\rightarrow c})$ represent the message from CN $c$ to VN $v$ (VN $v$ to CN $c$) for iteration $\ell$. $\mathcal{N}(x)$ denotes the neighbor nodes of the node $x$. The initial messages are set as ${\hat m}^{(0)}_{c\rightarrow v}=0$ and the channel log-likelihood ratio (LLR) of VN $v$ is denoted by $m_v^{\rm ch}$.
For $\ell={1,\ldots,\overline{\ell}}$, the NMS decoding algorithm \cite{nachmani2018deep} propagates the following messages

\begin{align}
        m^{(\ell)}_{v\rightarrow c} &= \overline{w}^{(\ell)}_v m_{v}^{\rm ch} + \sum_{c'\in \mathcal{N}(v) \setminus c}  {\hat m}^{(\ell - 1)}_{c'\rightarrow v} \label{Eq:MS_1}\\
    {\hat m}^{(\ell)}_{c\rightarrow v} &= {w}^{(\ell)}_{c\rightarrow v} \left( \prod_{v'\in \mathcal{N}(c) \setminus v} {\rm sgn} \left( m^{(\ell)}_{v' \rightarrow c} \right) \right)   \min_{v'\in \mathcal{N}(c) \setminus v} |m^{(\ell)}_{v'\rightarrow c}|.
    \label{Eq:MS_2}
\end{align}
We call $\overline{w}^{(\ell)}_v$ and ${w}^{(\ell)}_{c\rightarrow v}$ the channel weight (ChW) and the CN weight (CW), respectively.
Finally, output LLRs $m_{v}^{\rm o}$ are computed at the last iteration $\overline{\ell}$ by 
\begin{equation*}
m_{v}^{\rm o} = m_{v}^{\rm ch} + \sum_{c\in \mathcal{N}(v)} {\hat m}^{({\ell})}_{c\rightarrow v}.
\end{equation*}

 \begin{figure}[t]
\centering
\includegraphics[scale=0.35]{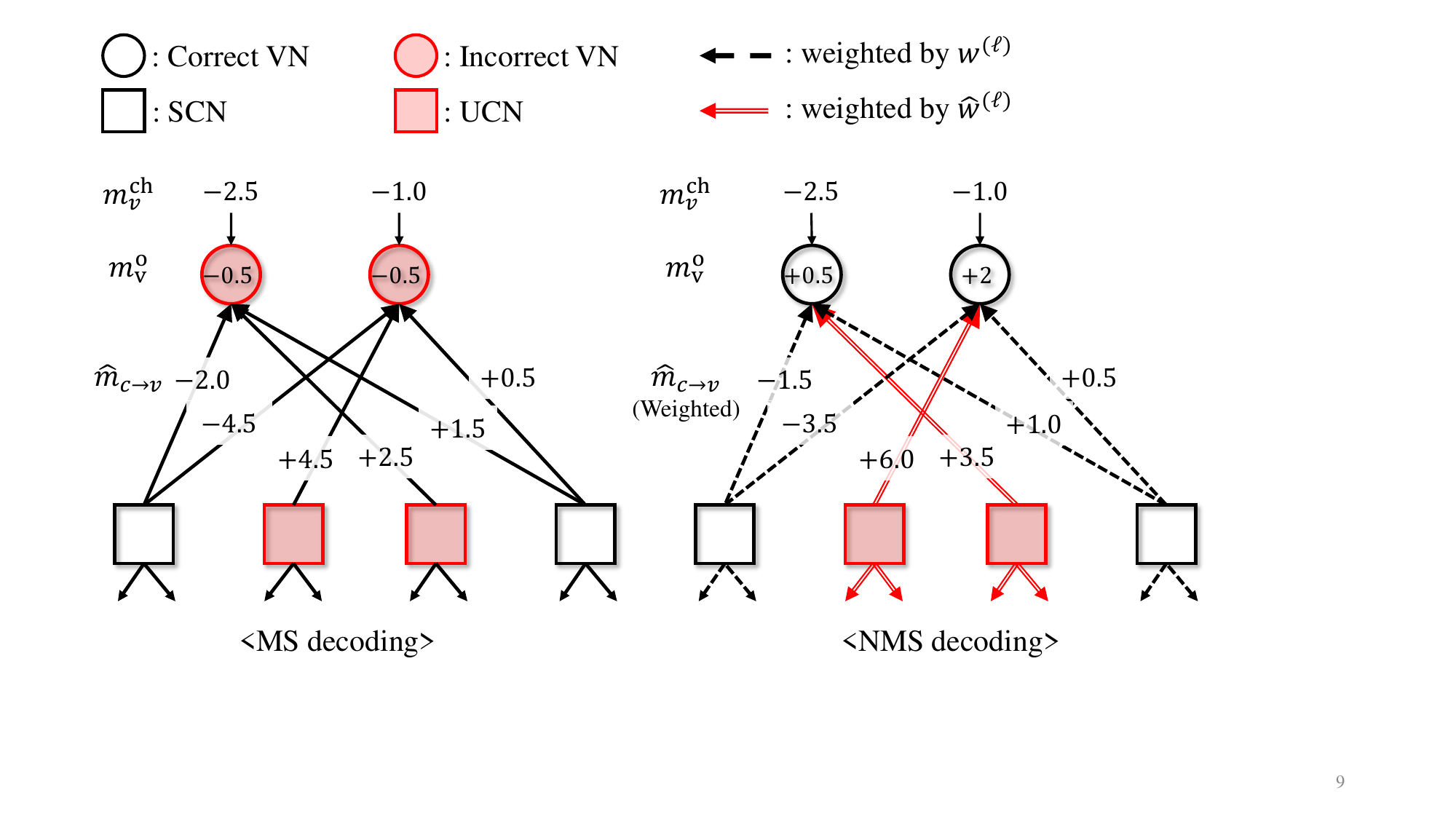}
\caption{An example involves introducing weights that enable successful decoding in a particular case.}
\label{Fig:Tanner_weights}
\end{figure}

The quantized NMS decoding algorithm is simply derived by quantizing the values of $m^{(\ell)}_{v\rightarrow c}$, ${\hat m}^{(\ell)}_{c\rightarrow v}$, and $m_v^{\rm ch}$. Quantized decoders are extensively employed in practical applications due to their reduced complexity and hardware-friendly nature. We employ a popular 5-bit uniform quantization setting with a maximum magnitude $7.5$ and a step size $\Delta=0.5$ to facilitate fair comparison with the existing error floor study
\cite{han2022deep}.

Fig.~\ref{Fig:Tanner_weights} illustrates an example of MS and NMS decoding, where LLR values $m_v^{\rm ch}$, ${\hat m}^{(\ell)}_{c\rightarrow v}$, and $m_v^{\rm o}$ are presented. Initially, two VNs are incorrect, but weighting ${\hat m}^{(\ell)}_{c\rightarrow v}$ corrects them. Assuming an all-zero codeword, negative values of $m_v^{\rm o}$ indicate erroneous VNs. For NMS decoding, ${\hat m}^{(\ell)}_{c\rightarrow v}$ on black dotted edges are weighted by $0.8$, while those on red double edges are weighted by $1.3$. This method for assigning weights to ${\hat m}^{(\ell)}_{c\rightarrow v}$ will be detailed in Subsection~\ref{SubSec:Dynamic}. The primary goal of this work is to optimize the weights of the NMS decoder to correct error patterns causing the error floor.

\subsection{Training Weights for the NMS Decoder}
 \begin{figure}[t]
\centering
\includegraphics[scale=0.43]{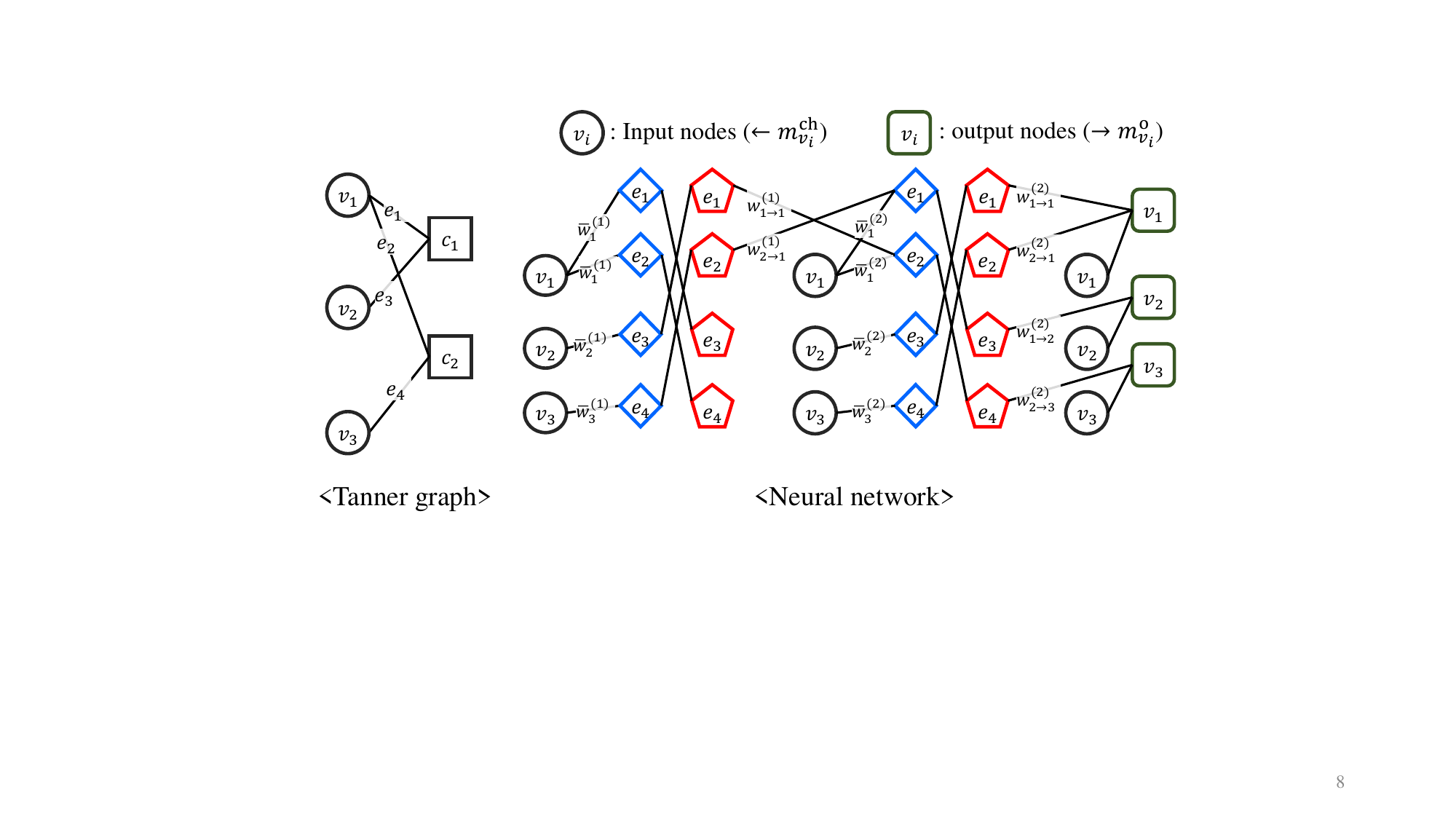}
\caption{A Tanner graph of an LDPC code and the corresponding neural network with a maximum iteration of $\overline{\ell}=2$.}
\label{Fig:Graph}
\end{figure}

Referring to (\ref{Eq:MS_1}) and (\ref{Eq:MS_2}), it is evident that when all weights are set to 1, the NMS decoding reduces to the MS decoding. Alternatively, by assigning $\overline{w}_v^{(\ell)}=1,w_{c\rightarrow v}^{(\ell)}=w$ for a single value of $w$, the resulting formula equals that of the WMS decoder. 
The diversity of weights in the NMS decoder enhances decoding performance over the WMS and MS decoders. However, assigning distinct weights to every edge, node, and iteration introduces excessive training complexity and memory requirements. To mitigate this, prior research has implemented weight sharing techniques, assigning the same value to weights having the same attributes. 
In this study, we primarily employ the protograph weight sharing technique \cite{dai2021learning}, which assigns the identical weight value to the $z$ edges (or nodes) from a proto edge (or node). The weight set is represented by $\{\overline{w}_{v_p}^{(\ell)}, w_{c_p \rightarrow v_p}^{(\ell)}\}_{\ell}$ for a proto VN $v_p$ and a proto CN ${c_p}$, with the total count of $(N+E)\overline{\ell}$. We refer to the weight set as the full weight set. 
Furthermore, applying temporal weight sharing \cite{nachmani2018deep}, which unifies weights across all iterations into a single value, results in the weight set $\{\overline{w}_{v_p}, w_{c_p \rightarrow v_p}\}$ with a total count of $(N+E)$. On the other hand, utilizing spatial weight sharing \mbox{\cite{lian2019learned, Andreev2021}}, where each iteration employs a single ChW and CW, simplifies the weight set to a form $\{\overline{w}^{(\ell)}, w^{(\ell)}\}_{\ell}$, with the total count adjusting to $2{\overline{\ell}}$.

Fig.~\ref{Fig:Graph} depicts a Tanner graph of an LDPC code and the corresponding neural network with $\overline{\ell}=2$. The neural network receives the channel LLR vector $(m_1^{\rm ch},\ldots, m_n^{\rm ch})$ as input and produces the output LLR vector $(m_1^{\rm o},\ldots, m_n^{\rm o})$.
For each decoding iteration, two hidden layers are allotted, and each layer contains one node (or neuron) corresponding to each edge in the Tanner graph. The VNs to CNs messages in (\mbox{\ref{Eq:MS_1}}) are generated by the odd hidden layers, whereas the even layers pass the CN to VN messages in (\mbox{\ref{Eq:MS_2})}.
The input layer connected to all odd hidden layers represents the incorporation of the channel LLR in (\ref{Eq:MS_1}). Messages passed from the $2\ell$-th to the $(2\ell+1)$-th hidden layer are weighted by CW ${w}^{(\ell)}_{c \rightarrow v}$, and those from the input layer to the $(2\ell-1)$-th hidden layer are weighted by ChW $\overline{w}^{(\ell)}_{v}$. For the loss function, several options exist, such as the BCE \cite{nachmani2018deep}, soft bit error rate (BER) \cite{lian2019learned}, and FER loss \mbox{\cite{Andreev2021, xiao2021faid}} functions. We implement the FER loss $\frac{1}{2}\left[ 1-{\rm sgn}\left({\rm min}~m_v^{\rm o} \right) \right]$ as our target is to minimize the FER in the error floor region.

To train the weights $\overline{w}^{(\ell)}_{v}$ and ${w}^{(\ell)}_{c \rightarrow v}$, we feed training samples into the neural network. Using channel modeling, such as the AWGN channel model, we generate a random received vector of length $n$ along with its corresponding channel LLR vector $(m_1^{\rm ch},\ldots, m_n^{\rm ch})$ to serve as a training sample. 
Received vectors are generated assuming the all-zero codeword. An important decision in training is to select the sampling points, or more specifically, the SNR region for generating received vectors. The SNR range within the waterfall region is typically selected to enhance waterfall performance. Therefore, a straightforward approach for improving error floor performance is to choose the SNR range within the error floor region. However, if training samples are collected from high SNRs within the error floor region, the loss function approaches zero, preventing weight updates. Consequently, a more sophisticated technique is required to train the NMS decoder with superior error floor performance.

\section{Proposed Training Methodology}\label{Sec:Proposed_Training}

\begin{figure}[t]
\centering
\includegraphics[scale=0.42]{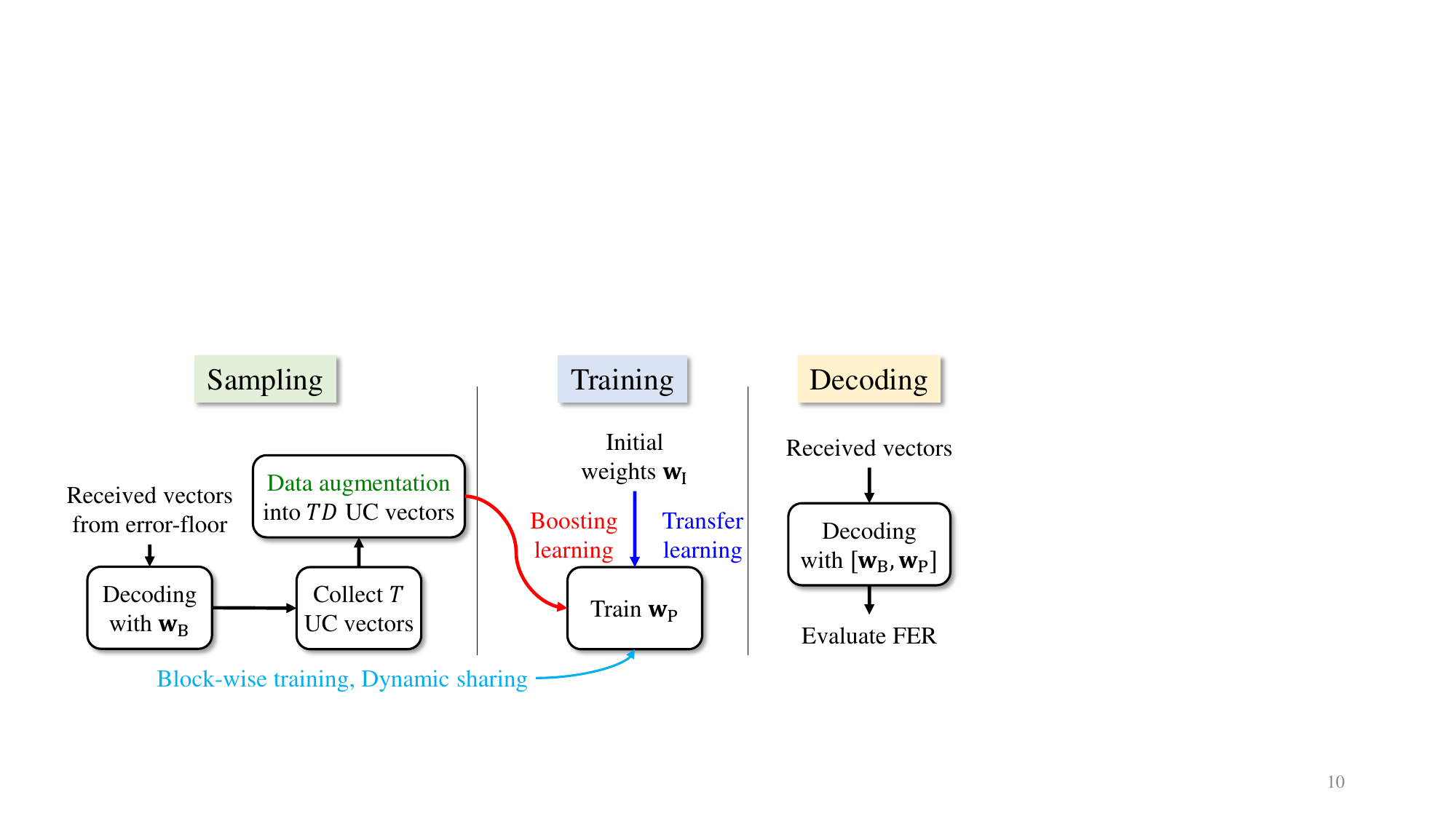}
\caption{Block diagram showing the proposed training methods.}
\label{Fig:Block_Diagram}
\end{figure}

In this section, we introduce our proposed training methodology across five subsections. The overall structure of the training methods is depicted in Fig.~\ref{Fig:Block_Diagram}.


\subsection{Boosting Learning Using Uncorrected Vectors}\label{SubSec:Boosting}

We divide the decoding process into two stages: i) the base decoding stage, covering iterations $\{1,\ldots,\ell_1\}$ with weights ${\bf w}_{\text{B}}$, and ii) the post decoding stage, covering iterations \mbox{$\{\ell_1+1,\ldots,\ell_1+\ell_2=\overline{\ell}\}$} with weights ${\bf w}_{\text{P}}$. Adopting the philosophy of boosting learning, the post decoder is designed to learn from the failure of the base decoder, thereby compensating for the weakness of the base decoder. Note that while the decoding process is conceptually divided into two stages, in practice, it is operated by a single decoder utilizing trained weights ${\bf w}_{\text{B}}$ and ${\bf w}_{\text{P}}$.
 
For the training of the base decoder, we follow the conventional training method \cite{nachmani2018deep}: the base decoder is trained with received vectors sampled at the waterfall region, ${\rm E}_{\rm b}/{\rm N}_0$ values $\{2.0,2.5,3.0,3.5,4.0\}$ (dB). Additionally, we employ the spatial weight sharing (i.e., ${\bf w}_{\text{B}}=\{\overline{w}^{(\ell)}, w^{(\ell)}\}_{\ell=1}^{\ell_{1}}$) because it shows a similar performance with the full weight set in the waterfall region. The weights are initially set to $1$, which means that the NMS decoder is equivalent to the MS decoder before training.

\begin{figure}[t]
\centering
\subfigure[]{\includegraphics[scale=0.38]{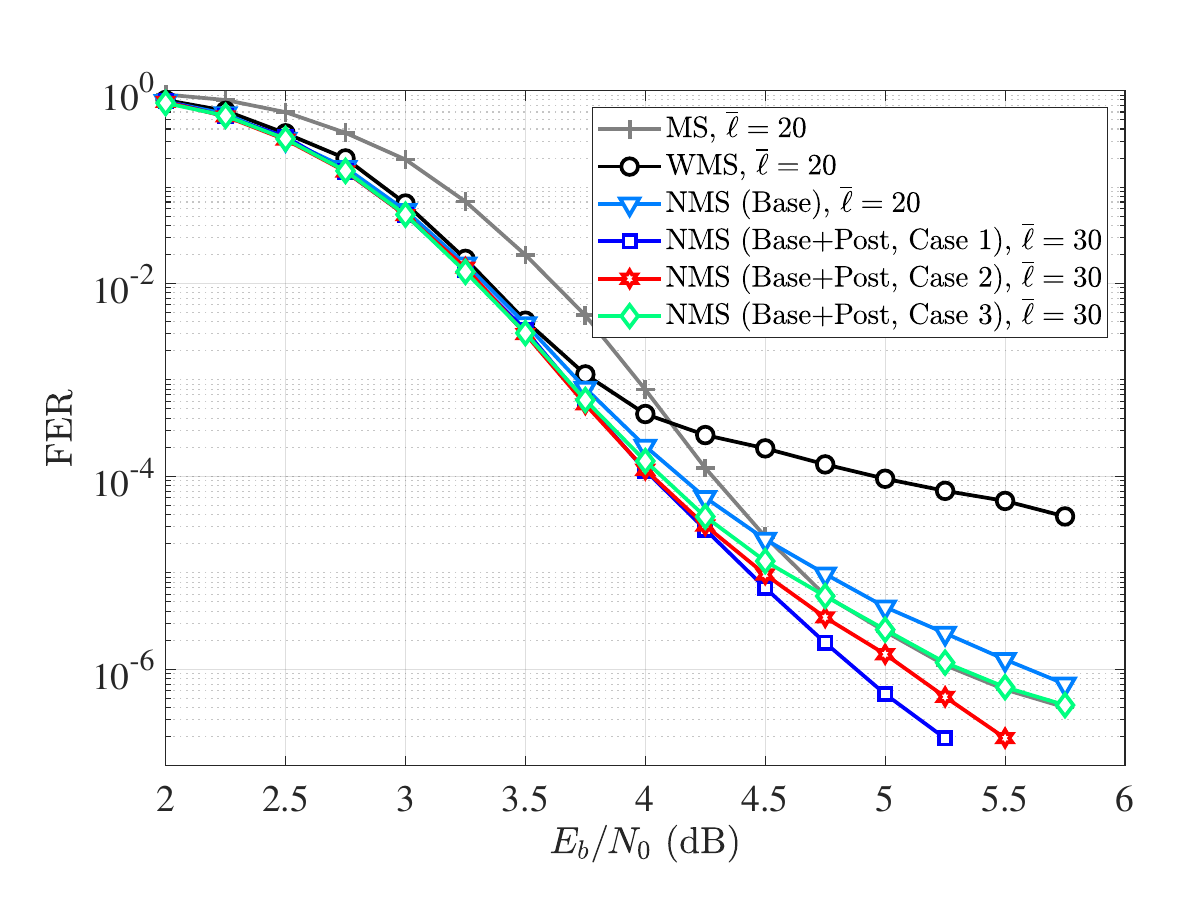}}
\subfigure[]{\includegraphics[scale=0.38]{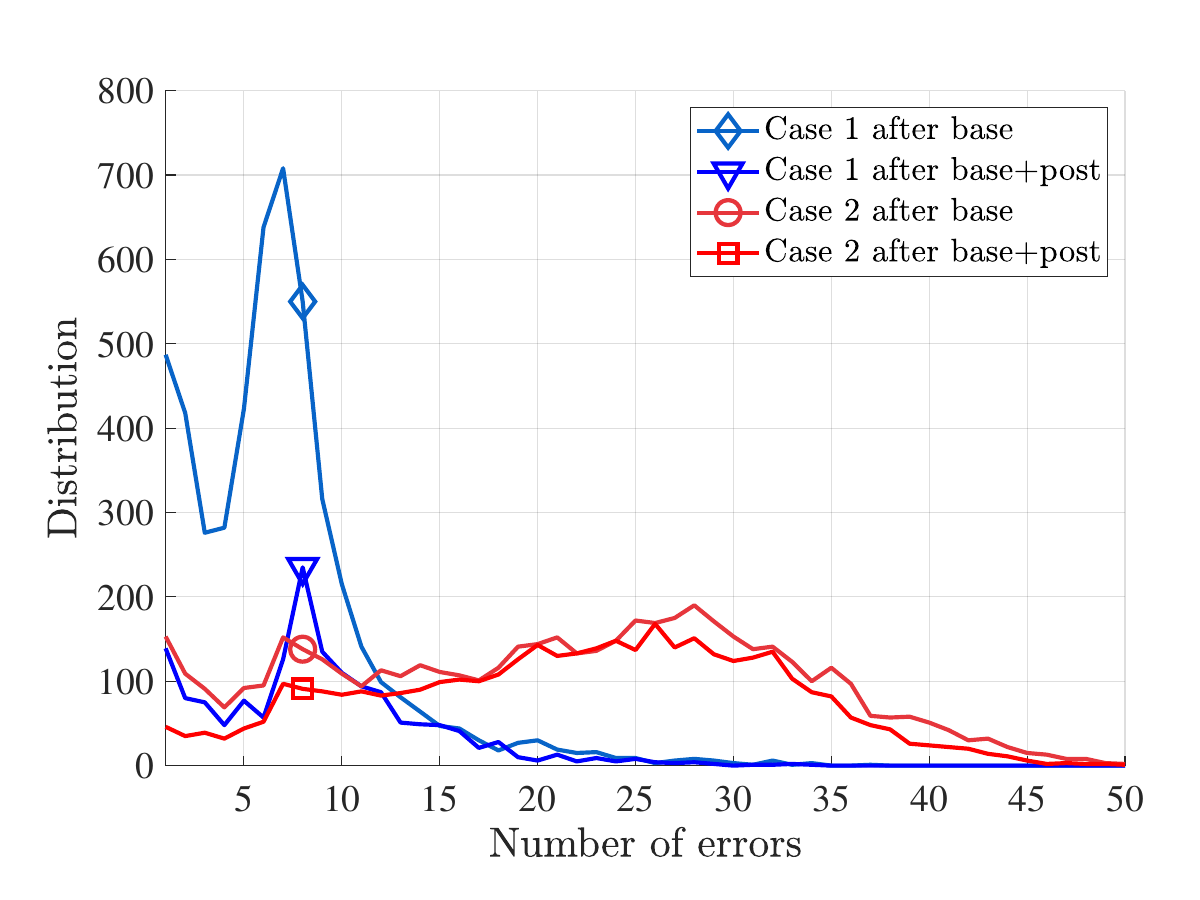}}
\caption{(a) FER performances of the MS, WMS, NMS decoders and (b) Error distributions after base and post decoding for Case 1 and Case 2.}
\label{Fig:FER_iter30}
\end{figure}

Fig.~\ref{Fig:FER_iter30}(a) compares the decoding performance with $\overline{\ell}=20$ for the base NMS, MS, and WMS decoders. The WMS decoder uses a single weight of $0.75$. While the WMS decoder shows superior waterfall performance to the MS decoder, it exhibits a severe error floor. The base NMS decoder, trained in the waterfall region, shows waterfall performance that is similar to that of the WMS decoder. 
However, its performance deteriorates compared to the MS decoder in the error floor region. To improve the error floor performance, one can extend the training region to include the error floor region. However, this approach is ineffective because of the almost zero FER loss in the error floor region. For example, the FER of the MS decoder (initial state of the NMS decoder) at $4.5$ dB is around $2\times 10^{-5}$, which implies that only a small fraction of training samples produce a non-zero loss. 

To effectively train the post decoder, we accumulate training samples by exclusively gathering UC vectors that the base decoder fails to correct. A UC vector consists of channel LLRs leading to a decoding failure. Subsequently, the post decoder is trained with the UC vectors. In terms of neural networks, the network is constructed up to $\overline{\ell}$, and the weights for iterations up to $\ell_1$ are fixed by the weight set ${\bf w}_{\text{B}}$ and the weight set ${\bf w}_{\text{P}}$ for iterations from $\ell_1+1$ to $\overline{\ell}$ is trained. For this experiment, we utilize the full weight set ${\bf w}_{\text{P}}=\{\overline{w}^{(\ell)}_{v_p}, w^{(\ell)}_{c_p \rightarrow v_p}\}_{\ell=\ell_1+1}^{\overline{\ell}}$ to observe the best performance, and we will apply weight sharing to ${\bf w}_{\text{P}}$ in the following subsection. 

To demonstrate the effectiveness of the proposed boosting learning technique, we compare the following three cases of training samples for the post decoder:
\begin{itemize}
    \item Case 1: UC vectors sampled at $4.5$ dB (error floor).
    \item Case 2: UC vectors sampled at $3.5$ dB (waterfall).
    \item Case 3: Received vectors sampled at $4.5$ dB. 
\end{itemize}

For Cases~1 and 2, we collect $60{,}000$ UC vectors, allocating $50{,}000$ for training, $5{,}000$ for validation, and $5{,}000$ for testing. Fig.~\ref{Fig:FER_iter30}(b) presents the distribution of the number of bit errors in decoded vectors after base and post decoding for the test samples. Observing Case~1 after base decoding, the majority of errors are concentrated in a small count, implying that most errors in the error floor region arise from small trapping sets. For convenience, we refer to vectors with $11$ or fewer errors as small-error vectors. In Case~1 after base+post decoding, it is evident that the post decoder successfully corrects most of the small-error vectors. The post decoder exhibits FER of $0.322$ among the $5{,}000$ test samples, denoted as test FER. This results in a reduction of FER from the base decoder to the cascaded decoder (base+post) by a factor of $0.322$ at $4.5$ dB, as shown in Fig.~\ref{Fig:FER_iter30}(a). The test FER will serve as an important performance metric throughout this paper.

\begin{table}[t]
\centering
\caption{Test BER and FER for representative loss functions.}
\begin{tabular}{c|ccc}
\hline
         & BCE \cite{nachmani2018deep}                  & Soft BER  \cite{lian2019learned}            & FER   \cite{xiao2021faid}               \\\hline\hline
Test BER & $5.6\times 10^{-3}$ & $4.7\times 10^{-3}$ & $5.0\times 10^{-3}$ \\
Test FER & $0.470$              & $0.379$              & $0.322$          \\\hline   
\end{tabular}
\label{Table:Loss}
\end{table}

In contrast, Case~2 after base decoding shows a wider distribution of error counts, as it collects UC vectors at the waterfall region. This distribution remains largely unchanged even after post decoding. Therefore, the strategy for Case 1 is more effective in reducing the error floor, as clearly shown in Fig.~\ref{Fig:FER_iter30}(a). In other words, it can be concluded that collecting training samples from the error floor region is vital for training decoders specialized in combating the error floor. 
The post decoder is trained to introduce a different form of diversity compared to
the base decoder, addressing
the error floor caused by a small number of errors.

In Case~3, the base decoding stage corrects most of the training samples and leaves very few errors for the post stage. This causes the loss function to approach zero, and the weights of the post decoder remain nearly unchanged from their initial value of $1$. As a result, Fig.~\ref{Fig:FER_iter30}(a) shows that the error floor performance of Case~3 is similar to that of the MS decoder.

Before proceeding to the next subsection, we compare the loss functions in Table~\ref{Table:Loss}, assuming the boosting learning with Case 1. The result shows that the FER loss function, as expected, exhibits superior test FER performance although the BER performance is worse than that of the soft BER loss function. Since the problem of the error floor region is the gradual decrease in the FER curve rather than the BER curve, we choose to utilize the FER loss function.

\subsection{Block-wise Training Schedule}\label{SubSec:Schedule}

\begin{figure}[t]
\centering
\subfigure[]{\includegraphics[scale=0.5]{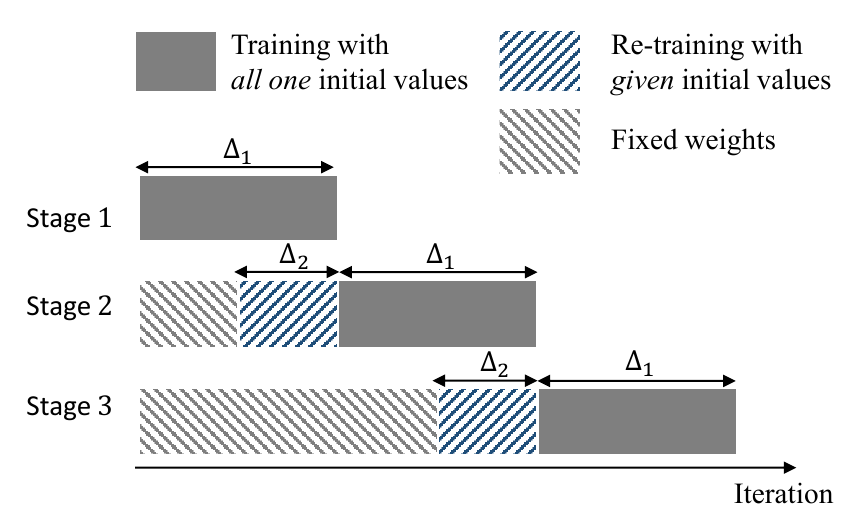}}
\subfigure[]{\includegraphics[scale=0.41]{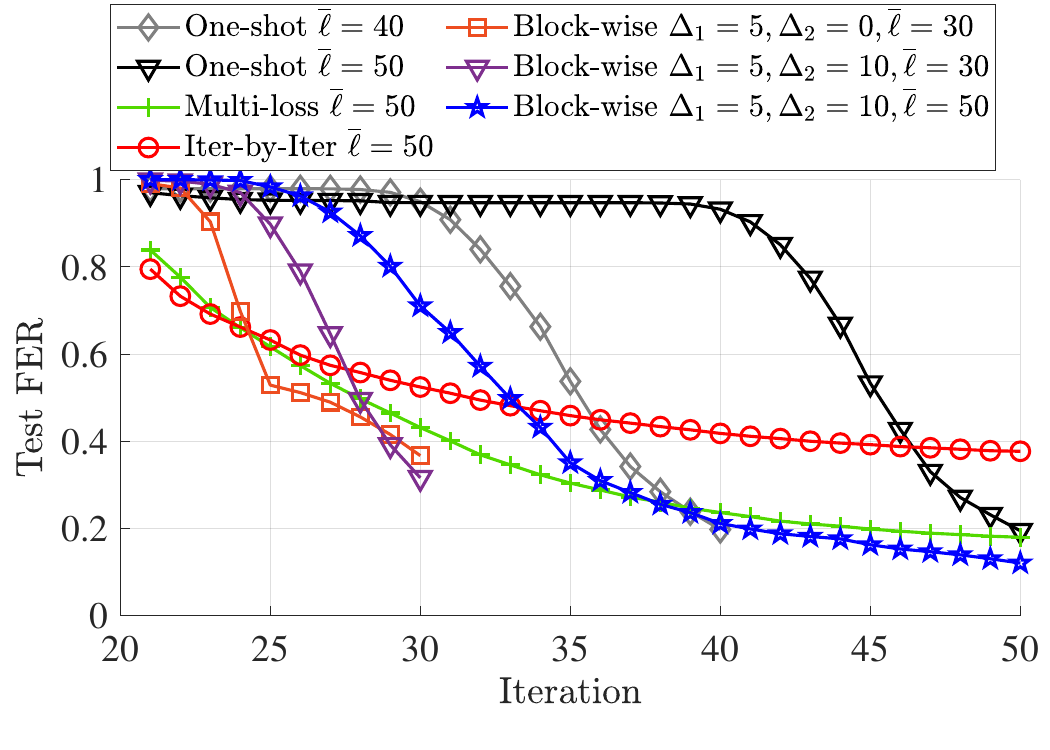}}
\caption{(a) Illustration of the block-wise training schedule and (b) comparison of the test FER across iterations.}
\label{Fig:Schedule}
\end{figure}

\begin{table}[t]
\centering
\caption{Test FER of the block-wise training with various $\Delta_1$ and $\Delta_2$.}
\begin{tabular}{c|ccccc}
\hline
$\Delta_1$ $\backslash$ $\Delta_2$ & $0$     & $5$     & $10$          & $15$    & $30$    \\\hline \hline
$1$                   & $0.376$ & $0.144$ & $0.123$       & $0.122$ & $0.150$  \\
$5$                   & $0.193$ & $0.134$ & ${\bf 0.112}$ & $0.127$ & $0.137$ \\
$10$                  & $0.142$ & $0.142$ & $0.13$        & $0.128$ & $0.13$  \\
$30$                  & $0.179$ &         &               &         &        \\
\hline
\end{tabular}
\label{Table:Schedule}
\end{table}

To further enhance error floor performance, we consider a larger number of iterations, $(\ell_2=30, \overline{\ell}=50)$, compared to $(\ell_2=10, \overline{\ell}=30)$ used in the previous subsection. However, this induces deep hidden layers which, in turn, exacerbates the vanishing gradient problem. To address this problem, we propose a block-wise training schedule as illustrated in Fig.~\ref{Fig:Schedule}(a). 
The training schedule incrementally trains weights associated with $\Delta_1$ iterations per stage. Initially, weights for the first $\Delta_1$ iterations are trained, and then the focus shifts to the next $\Delta_1$ iterations. The weights of previous $\Delta_2$ iterations are retrained using the trained weights from earlier stages as initial weights. 
This retraining is introduced to help the training process avoid local minima. Note that the one-shot training \cite{nachmani2018deep} belongs to a special case of $\Delta_1=\ell_2, \Delta_2=0$, and the iter-by-iter method \cite{dai2021learning} is of \mbox{$\Delta_1=1$}, $\Delta_2=0$.

To optimize the block-wise training parameters $\Delta_1$ and $\Delta_2$, we evaluate test FER for varying configurations of these parameters as shown in Table \ref{Table:Schedule}. For scenarios without retraining (i.e., $\Delta_2=0$), increasing $\Delta_1$ from $1$ to $10$ achieves performance gains. 
This can be attributed to the training of more weights simultaneously, aiding in escaping local minima. 
Nonetheless, excessively large values of $\Delta_1$ can result in undertraining of the weights due to the vanishing gradient problem, thereby degrading performance. 
Table \ref{Table:Schedule} indicates that the optimal result is obtained with $\Delta_1=5$ and $\Delta_2=10$.

Fig.~\ref{Fig:Schedule}(b) shows the evolution of test FER according to iteration numbers. 
For the one-shot training \cite{nachmani2018deep}, the vanishing gradient problem impedes the effective training of early iteration weights. This results in consistent test FER up to iteration $40$ for $\overline{\ell}=50$. Since this behavior is observed for $\overline{\ell}=40$ as well, the test FER values for $\overline{\ell}=40$ and $\overline{\ell}=50$ are similar.
The iter-by-iter method \cite{dai2021learning} optimizes weights for each iteration one-by-one, reducing test FER from the initial iteration. However, this approach gets trapped in a local minimum and leads to only slow FER improvements over iterations. A similar trend is observed with the multi-loss method \cite{nachmani2018deep}.

For the block-wise training schedule, we first show the cases of $(\Delta_1=5, \Delta_2=0)$ and $(\Delta_1=5, \Delta_2=10)$ for $\overline{\ell}=30$. The parameter set $(\Delta_1=5, \Delta_2=0)$ exhibits a superior test FER at iteration $25$ compared to the iter-by-iter method due to the simultaneous training of multiple iterations. The introduction of retraining ($\Delta_1=5$, $\Delta_2=10$) shows a worse result at iteration $25$ compared to $(\Delta_1=5, \Delta_2=0)$, but further improves performance at the last iteration $30$. This means that retraining adjusts
the weights of intermediate iterations to lead to enhanced results in the last iteration.
For $\overline{\ell}=50$, the block-wise training schedule with ($\Delta_1=5$, $\Delta_2=10$) achieves the lowest test FER, outperforming other methods: $0.11$ for the block-wise, $0.16$ for the multi-loss \cite{nachmani2018deep}, $0.18$ for the one-shot \cite{nachmani2018deep}, $0.37$ for the iter-by-iter \cite{dai2021learning}.

\subsection{Dynamic Weight Sharing}
\label{SubSec:Dynamic}

We propose a new weight sharing technique that dynamically allocates weights during the decoding process. This method assigns distinct weights to UCNs and SCNs, as illustrated in Fig. \ref{Fig:Tanner_weights}.
It utilizes the decoder's ability to distinguish between SCNs and UCNs via parity check equations.
In addition, we incorporate spatial weight sharing. For iteration $\ell$, the decoder verifies the parity check equation for each CN and assigns the CW $w^{(\ell)}$ to SCNs and the UCN weight (UCW) $\hat{w}^{(\ell)}$ to UCNs. The assignment of weights is dynamic, changing with each iteration. The weight set is represented by $\{\overline{w}^{(\ell)}, w^{(\ell)},{\hat {w}}^{(\ell)}\}_{\ell}$. The total number of distinct weights for the post decoder becomes $3\ell_2$.

\begin{figure}[t]
\centering
\includegraphics[scale=0.4]{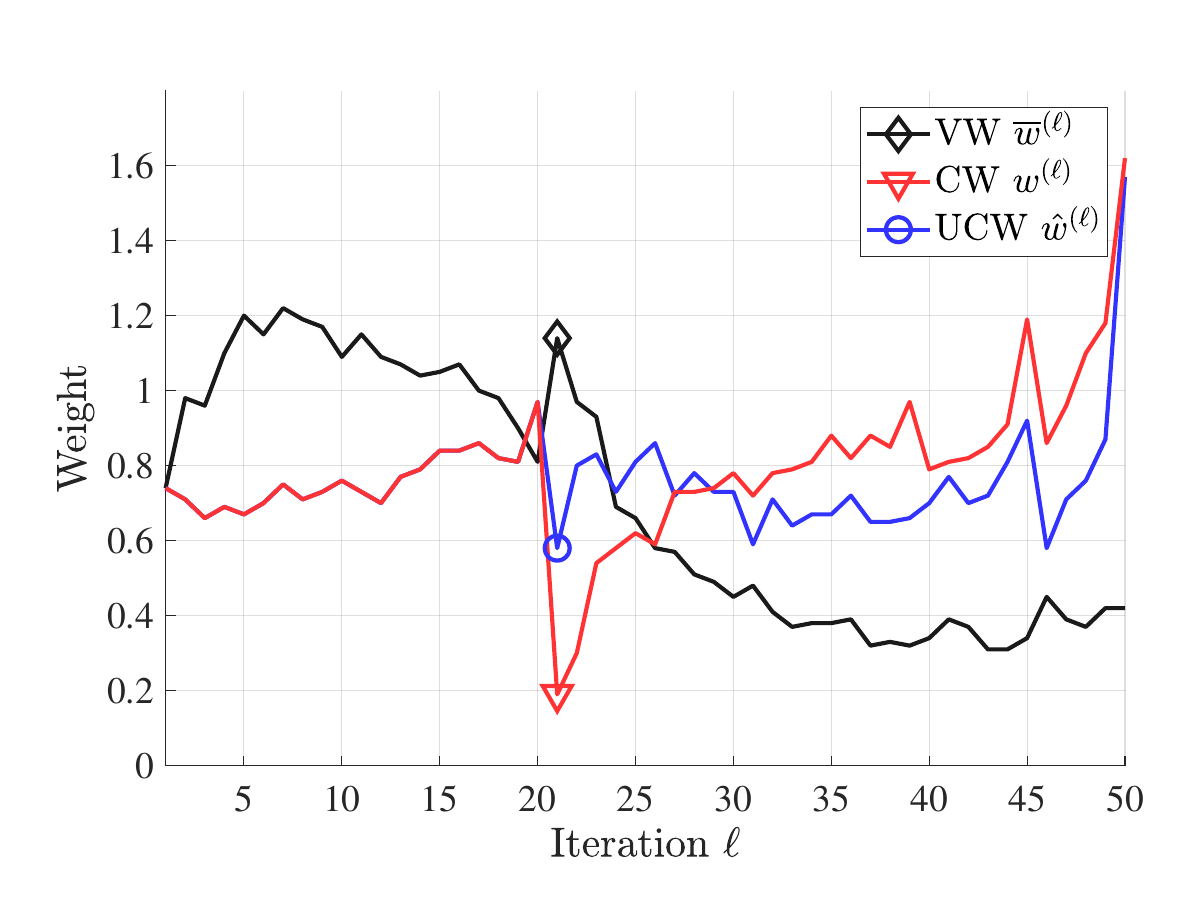}
\caption{Evolution of the trained weights according to iterations.}
\label{Fig:Weights}
\end{figure}

\begin{table}[t]
\centering
\caption{Test FER for various weight sharing techniques.}
\begin{tabular}{c|cc}
\hline
                 & Test FER & Number of weights    \\ \hline\hline
Full $\{\overline{w}_{v_p}^{(\ell)}, w_{c_p \rightarrow v_p}^{(\ell)}\}_{\ell}$   & $0.112$  & $(N+E)\ell_{2}=3360$ \\
Spatial $\{\overline{w}_{v_p}, w_{c_p \rightarrow v_p}\}$  & $0.168$  & $2\ell_{2}=60$       \\
Temporal $\{\overline{w}^{(\ell)}, w^{(\ell)}\}_{\ell}$ & $0.186$  & $(N+E)=112$          \\
Dynamic $\{\overline{w}^{(\ell)}, w^{(\ell)},{\hat {w}}^{(\ell)}\}_{\ell}$ & $0.111$  & $3\ell_{2}=90$    \\  \hline
\end{tabular}
\label{Table:UCN_weights}
\end{table}

Given the implementation of our techniques so far, including training with UC vectors sampled at ${\rm E}_{\rm b}/{\rm N}_0=4.5$ dB and utilizing the block-wise training schedule with $\Delta_1=5,\Delta_2=10$, we compare different weight sharing methods in Table.~\ref{Table:UCN_weights}. The proposed dynamic sharing method attains performance comparable to that achieved with the full weight set while the other methods lead to performance degradation. In addition, our method achieves a \SI{97}{\%} reduction in the total number of weights.

Fig.~\ref{Fig:Weights} shows the trained weights using our sharing technique. With only three distinct weights per iteration -- ChW $\overline{w}^{(\ell)}$, CW $w^{(\ell)}$, and UCW $\hat{w}^{(\ell)}$-- the weight evolution can be depicted simply in a two-dimensional graph. In the base stage, the CW and ChW values are roughly around $0.75$ and $1$, respectively, which are the weights of the WMS decoder. As a result, the WMS and NMS decoders perform similarly in the waterfall region.
On the contrary, a significant change in weights is observed in the post stage. At iteration $21$, the ChW increases and CW decreases substantially. This means that the channel LLR values are given more weight when performing the sum operation (\ref{Eq:MS_1}), while the messages coming from CNs are attenuated. From iteration $22$ onward, the CW gradually increases and the ChW decreases, suggesting an increasing dependence on CN messages. This dynamic weight evolution enhances decoding diversity in the post decoding stage that lowers the error floor.

The method differing UCW and CW has been explored in prior works \cite{shah2021neural, wu2010adaptive}. However, the method in \cite{wu2010adaptive} employs a single suppression factor $\rho$ to modify the UCW based on the CW (i.e., ${\hat {w}}^{(\ell)} = (1+\rho) w^{(\ell)}$), limiting the capability to manage various error patterns effectively. Meanwhile, the scheme in \cite{shah2021neural} assigns the UCW to all $z$ CNs derived from a single proto CN if any one of them is unsatisfied. This is not suitable for correcting small-error vectors with few UCNs since it does not consider individual CNs distinctively, whereas we pinpoint specific UCNs and apply weights individually.

\subsection{Ablation Study}

\begin{table}[t]
\centering	
 \caption{Ablation study for the $(n=576, k=432)$ WiMAX LDPC code.}
{\footnotesize
\begin{tabular}{c|c|c|c}
Learning & Schedule & Sharing &\\\hline
O: Boosting & O: Block-wise & O: Dynamic &  FER \\ 
X: Conv. &  X: One-shot & X: No sharing & (${\rm E}_{\rm b}/{\rm N}_0$ 5.0dB) \\ \hline
X                     & X                         & X                                      & $3.14\times 10^{-6}$ \\
O                     & X                         & X                                      & $2.77\times 10^{-7}$ \\
O                     & O                         & X                                      & $1.84\times 10^{-7}$ \\
O                     & O                         & O                                      & $1.85\times 10^{-7}$
\end{tabular}
}
\label{Table:Ablation}
\end{table}

The ablation study for the techniques introduced so far is presented in Table \mbox{\ref{Table:Ablation}}. For the $(n=576, k=432)$ WiMAX LDPC code, using boosting learning lowers FER at ${\rm E}_{\rm b}/{\rm N}_0=5.0$ dB by more than one order of magnitude, and block-wise training reduces it by approximately $1.5$ times. Additionally, dynamic sharing significantly reduces the number of weight sets with only minimal degradation in performance. In summary, the most essential technique for improving performance among the proposed methods is boosting learning.

\subsection{Transfer Learning}

 \begin{figure}[t]
\centering

\includegraphics[scale=0.51]{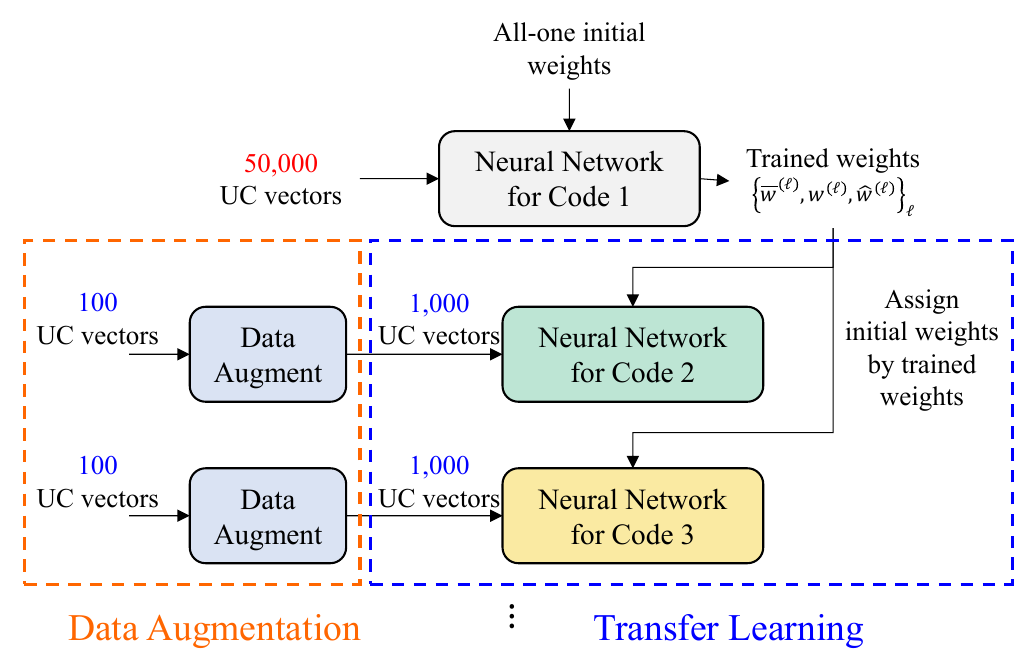}

\caption{Illustration of transfer learning and data augmentation.}
\label{Fig:Transfer_Learning}
\end{figure}

While conventional learning methods using received vectors as training samples have an immediate sampling process, the proposed boosting learning requires numerous decoding trials to collect UC vectors, leading to a time-intensive sampling process. For example, obtaining $T$ UC vectors requires $T$/$\rm FER$ decoding trials on average. 
In previous sections, we collect $50{,}000$ UC vectors at the base decoding FER of $2\times10^{-5}$. This sampling process requires $2.5\times 10^{9}$ decoding trials on average. If we require extremely low FER and training of multiple codes, the sampling process may become infeasible.

Transfer learning, a popular technique in machine learning, offers a solution to this issue. Consider two different codes, Code 1 and Code 2. As illustrated in Fig. \ref{Fig:Transfer_Learning}, the weights trained for Code 1 serve as the initial weights for training Code 2. 
However, the transfer learning method in the machine learning domain, which typically involves a fixed network and fine-tuning of weights for a different task, is not directly applicable for NMS decoders due to their varying neural network structures depending on underlying codes. If the full weight set is used, two codes must have the same number of edges and nodes to enable one-to-one weight transfer. 
On the contrary, the spatial and dynamic weight sharing techniques enable the application of transfer learning across diverse code structures.

 \begin{figure}[t]
\centering

\includegraphics[width=0.9\linewidth]{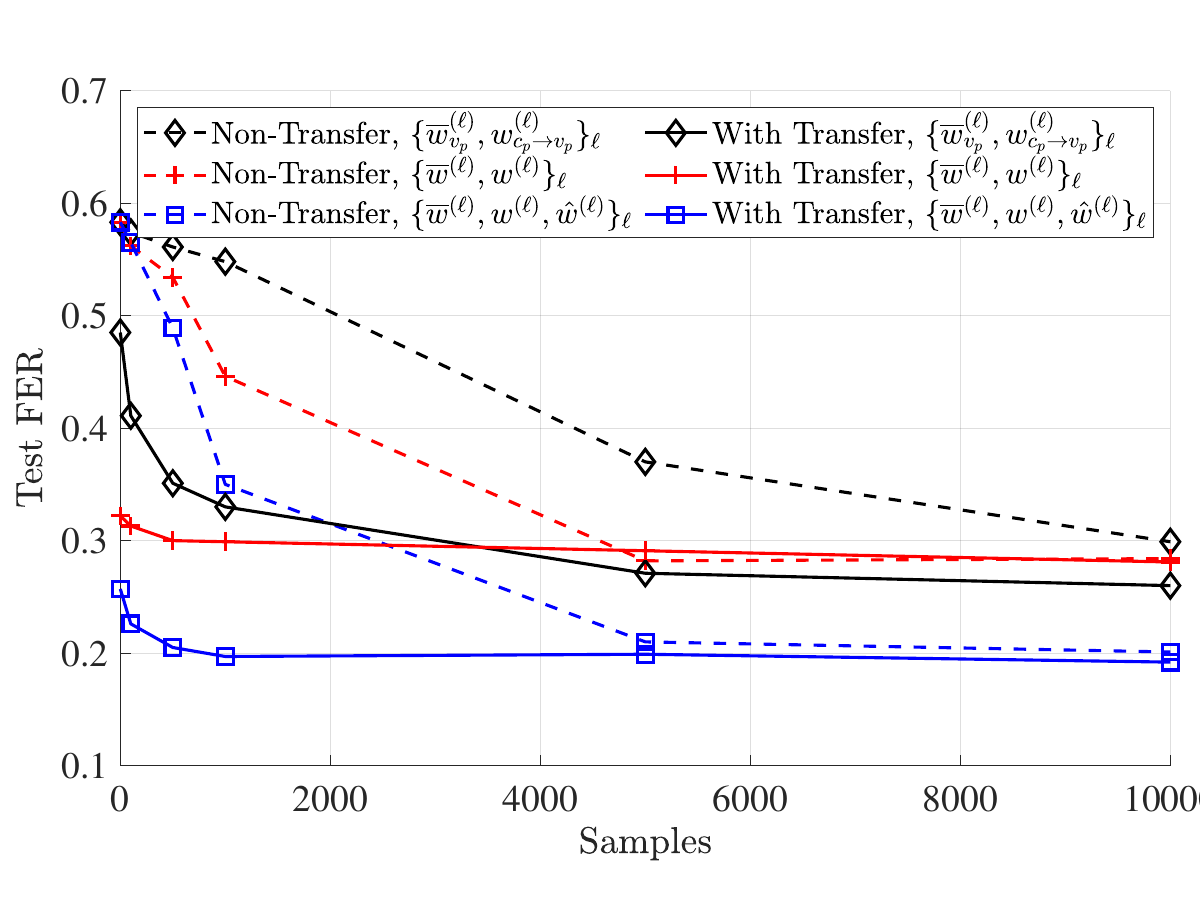}

\caption{Comparison of the test FER according to the number of training samples.}
\label{Fig:Transfer_Samples}
\end{figure}

Fig.~\ref{Fig:Transfer_Samples} shows the test FER according to the number of training samples used for the post decoder with $15$ iterations and the WiMAX LDPC code. Without transfer learning and using the full weight set $\{\overline{w}_{v_p}^{(\ell)}, w_{c_p \rightarrow v_p}^{(\ell)}\}_{\ell}$ (black dotted lines), the test FER begins at $0.56$ with zero training sample (i.e., no training) and reduces to $0.30$ after using $10{,}000$ samples. Though not shown in the figure, the test FER stabilizes at $0.21$ with $50{,}000$ samples. 
Using shared weights, $\{\overline{w}^{(\ell)}, w^{(\ell)}\}_{\ell}$ for spatial sharing and $\{\overline{w}^{(\ell)}, w^{(\ell)},{\hat {w}}^{(\ell)}\}_{\ell}$ for dynamic sharing, the test FER is saturated at their maximum performances with $10{,}000$ samples. This shows another advantage of weight sharing: Fewer trainable weights lead to fewer required sample counts.

\begin{table}[t]
\centering
\caption{Test FER for transfer learning with various variations}
\begin{tabular}{c|cccc}

    Variation                 & Code                     & Rate                     & Channel                  & Decoder                  \\\hline
From                           & WiFi                     & R1/2                     & Rayleigh                 & 4bit                     \\
To                             & WiMAX                    & R3/4                     & AWGN                     & 5bit                     \\ \hline\hline
Transfer with $0$ sample              & \multirow{2}{*}{$0.257$} & \multirow{2}{*}{$0.292$} & \multirow{2}{*}{$0.197$} & \multirow{2}{*}{$0.236$} \\
 (No training)      &                          &                          &                          &                          \\ \hline
Transfer with $1{,}000$ samples              & \multirow{2}{*}{$0.197$} & \multirow{2}{*}{$0.212$} & \multirow{2}{*}{$0.194$} & \multirow{2}{*}{$0.192$} \\
(Fine tunning) &                          &                          &                          &                          \\ \hline
\end{tabular}
\label{Table:Transfer_var}
\end{table}

For transfer learning, the initial weights for the WiMAX LDPC code of $M=6, N=24, E=88, R=3/4$ are set by the pre-trained weights for the WiFi LDPC code $M=4, N=24, E=88, R=5/6$. Since both the WiFi and WiMAX codes have the same number of edges $E$ and nodes $N$, a direct transfer is available without weight sharing. Nonetheless, as shown in Fig.~\ref{Fig:Transfer_Samples}, the performance of the full weight set converges at $10{,}000$ samples (black solid line), while using weight sharing significantly reduces the required number of samples to $1{,}000$ (blue and red solid lines). 
Notably, the initial test FER with weight sharing is already low even before training. A small amount of fine-tuning is enough to obtain the optimal weight set. As a result, the combination of transfer learning and weight sharing reduces the required number of samples from $50{,}000$ to $1{,}000$, thereby accelerating the sampling process by a factor of $50$.

Table \ref{Table:Transfer_var} presents a range of experiments to show the effectiveness of transfer learning across different code types, code rates, channels, and decoders. The target conditions for transfer learning are the WiMAX LDPC code with code rate of $3/4$, the AWGN channel, and the 5-bit quantized decoder, which have been assumed as a running example so far. Among them, we modify one condition at a time and derive the pre-trained weights. For the code type, we use the pre-trained weights from the WiFi LDPC code, whose results are also in Fig. \ref{Fig:Transfer_Samples}. For all cases in Table \ref{Table:Transfer_var}, the fine tuning with $1{,}000$ samples improves the test FER compared to the baseline FER with merely transferring of weights, showing the effectiveness of transfer learning. 

In addition, these results demonstrate the flexibility of the proposed boosting learning method, which is applicable to various code types, channels, and decoders (i.e., WiFi LDPC code, Rayleigh channel, low-precision decoder). In the remainder of the paper, we will further explore its applicability to 5G LDPC codes, short length LDPC codes, alternative quantization schemes, and the BP decoder. Our method can also be extended to broader classes, such as layered decoding \cite{shah2021neural} and optimized quantization schemes \cite{Buchberger2021}, which are left for future work.

\subsection{Data Augmentation}

\begin{table*}[t]
\centering
\caption{Comparison between the plain sampling method and the accelerated sampling method using transfer learning and data augmentation}
\begin{tabular}{c|c|cc|cc|cc|c}
\hline
\multirow{2}{*}{Weight set}                                                         & \multirow{2}{*}{Learning}   & \multicolumn{2}{c|}{Plain Sampling}  & \multicolumn{2}{c|}{Data Augmentation}  & \multicolumn{2}{c|}{Decoding Trials} & Training \\
                                                                                    &                             & UC vectors ($a$) & FER ($b$)          & UC vectors ($c$) & FER ($d$)             & ($a/b+c/d$)           & Gain          & Test FER \\ \hline\hline
$\{\overline{w}_{v_p}^{(\ell)}, w^{(\ell)}_{c_p \rightarrow v_p}\}_{\ell}$                 & Non-transfer                & $50000$           & $2\times10^{-5}$ & N/A               & N/A                 & $2.5\times 10^9$    & $1$           & 0.209    \\ \hline
\multirow{5}{*}{$\{\overline{w}^{(\ell)}, w^{(\ell)},{\hat {w}}^{(\ell)}\}_{\ell}$} & Non-transfer                & $10000$           & $2\times10^{-5}$ & N/A               & N/A                 & $5\times 10^8$      & $5$           & 0.201    \\ \cline{2-9} 
                                                                                    & Transfer                    & $1000$            & $2\times10^{-5}$ & N/A               & N/A                 & $5\times 10^7$      & $50$          & 0.197    \\ \cline{2-9} 
                                                                                    & Transfer+Aug. ($\beta=0.5$) & $100$             & $2\times10^{-5}$ & $1000$            & $1.7\times 10^{-4}$ & $1.09\times 10^7$   & $230$         & 0.197    \\ \cline{2-9} 
                                                                                    & Transfer+Aug. ($\beta=0.7$) & $100$             & $2\times10^{-5}$ & $1000$            & $5.2\times 10^{-4}$ & $6.92\times 10^6$   & ${\bf 361}$   & 0.199    \\ \cline{2-9} 
                                                                                    & Transfer+Aug. ($\beta=0.9$) & $100$             & $2\times10^{-5}$ & $1000$            & $8.6\times 10^{-4}$ & $6.16\times 10^6$   & $405$         & 0.213    \\ \hline
\end{tabular}

\label{Table:Augmentation}
\end{table*}

In this subsection, we introduce the use of importance sampling for data augmentation to further speed up the sampling process. Importance sampling was used in estimating error floor performance of LDPC codes \cite{Richardson2003, Dolecek2009}. Under the assumption of the binary phase shift keying (BPSK) modulation and all-zero codeword, stronger noise to variable nodes (VNs) within the trapping set is introduced by subtracting $\beta$ from their received values $y_i$. (i.e., $y_i = 1 -\beta + n_{\rm ch}$ for $i$ within the trapping set, where $n_{\rm ch}$ is the channel noise.) This approach allows for the rapid evaluation of relatively high FER, which is used to estimate the actual FER. Thus, we can estimate extremely low actual FER with a small number of decoding trials.

The proposed data augmentation method also generates intensified noise at specific bit positions. As depicted in Fig.~\ref{Fig:Transfer_Learning}, the data augmentation block processes $T$ input UC vectors to produce $T\times D$ augmented UC vectors, where $T=100$ and $T\times D=1{,}000$. For each input UC vector, the base decoding is performed to identify the positions of error bits. These error bit positions, denoted as $\mathcal E$, are identified at the iteration when the number of UCNs is minimal. Subsequently, we accumulate $D$ UC vectors using the base decoder under a modified channel model where received vectors are generated from the AWGN channel as usual, and received values for VNs in $\mathcal E$ are shifted by $-\beta$. This shift increases the probability of decoding failures and enables the rapid collection of $D$ additional UC vectors that share characteristics with the input UC vector. Each UC vector entering the data augmentation block is augmented into $D$ UC vectors, thereby generating a total of $T\times D$ UC vectors from $T$ input UC vectors. Note that the error bit positions do not necessarily form a trapping set, as they are identified through the actual decoding process. Since we do not rely on algorithms to find trapping sets, data augmentation is feasible for long codes and is fast enough to collect augmented UC vectors.

Table \ref{Table:Augmentation} shows how transfer learning and data augmentation accelerates the sampling process. Without weight sharing, collecting $50{,}000$ UC vectors (indicated in ($a$)) with FER of the base decoder (indicated in FER ($b$)) requires $a/b=2.5\times 10^{9}$ decoding trials using the plain sampling method. 
By applying the dynamic weight sharing technique $\{\overline{w}^{(\ell)}, w^{(\ell)},{\hat {w}}^{(\ell)}\}_{\ell}$, the required number of UC vectors is reduced to $10{,}000$ without transfer learning and further to $1{,}000$ with the addition of transfer learning. For data augmentation, Table \ref{Table:Augmentation} presents results for $T=100$, $D=10$, and various $\beta$ values. With data augmentation, the augmented UC vectors ($c$) equal $T\times D=1{,}000$, and FER ($d$) represents FER under the modified channel with intensified noise. The decoding trials in data augmentation are $c/d$. Thus, the total number of decoding trials is given by $a/b+c/d$. The weight sharing technique alone reduces the required trials by a factor of $5$ and accelerates the sampling speed by the same factor. Transfer learning adds a 10-fold increase in speed, and this gain is further increased by data augmentation. As $\beta$ increases, FER ($d$) in data augmentation increases, giving more gains. However, augmented samples with high $\beta$ might differ in characteristics from those collected through plain sampling, which leads to a degraded test FER performance. Therefore, we opt for $\beta=0.7$, achieving a $\times 361$ increase in sampling speed without significant loss in the test FER.
Boosting learning accompanied by transfer learning and data augmentation would be highly useful, especially in environments that need to support numerous code rates and lengths, such as 5G LDPC codes.

Additionally, to eliminate the need for validation samples, we adopt the step decaying learning rate strategy. Until now, $5{,}000$ UC vectors have been allocated for validation samples. However, since the NMS decoder is not prone to overfitting on training data, the weights selected by validation samples have similar performance to those at the last epoch. To further enhance stability in convergence, we use the step decaying learning rate, starting from $0.001$ and halving it every $20$ epochs, and select the weights at the last epoch without validation samples.

\section{Performance Evaluation}\label{Sec:Evaluation}
In this section, we apply the proposed training methods to design the boosted NMS decoder and evaluate its performance with standard LDPC codes in various applications, including WiMAX, WiFi, and 5G NR systems.

\subsection{Evaluation Environment}
Training is carried out using the Adam optimizer \cite{kingma2014adam} with TensorFlow, on an NVIDIA GeForce RTX A5000 GPU and an AMD EPYC 7763 CPU. For fast simulation during the sampling process and FER evaluation, the C programming language is employed. To guarantee the uniqueness of each random experiment, a long-cycle noise generator, specifically the Mersenne Twister algorithm \cite{Matsumoto1998}, is utilized. The FER performances are evaluated using the Monte Carlo method, with experiments repeated until a minimum of $100$ UC vectors are identified. 

\subsection{Result for WiMAX and WiFi LDPC Codes}

\begin{figure*}[t]
\centering
\subfigure[]{\includegraphics[scale=0.4]{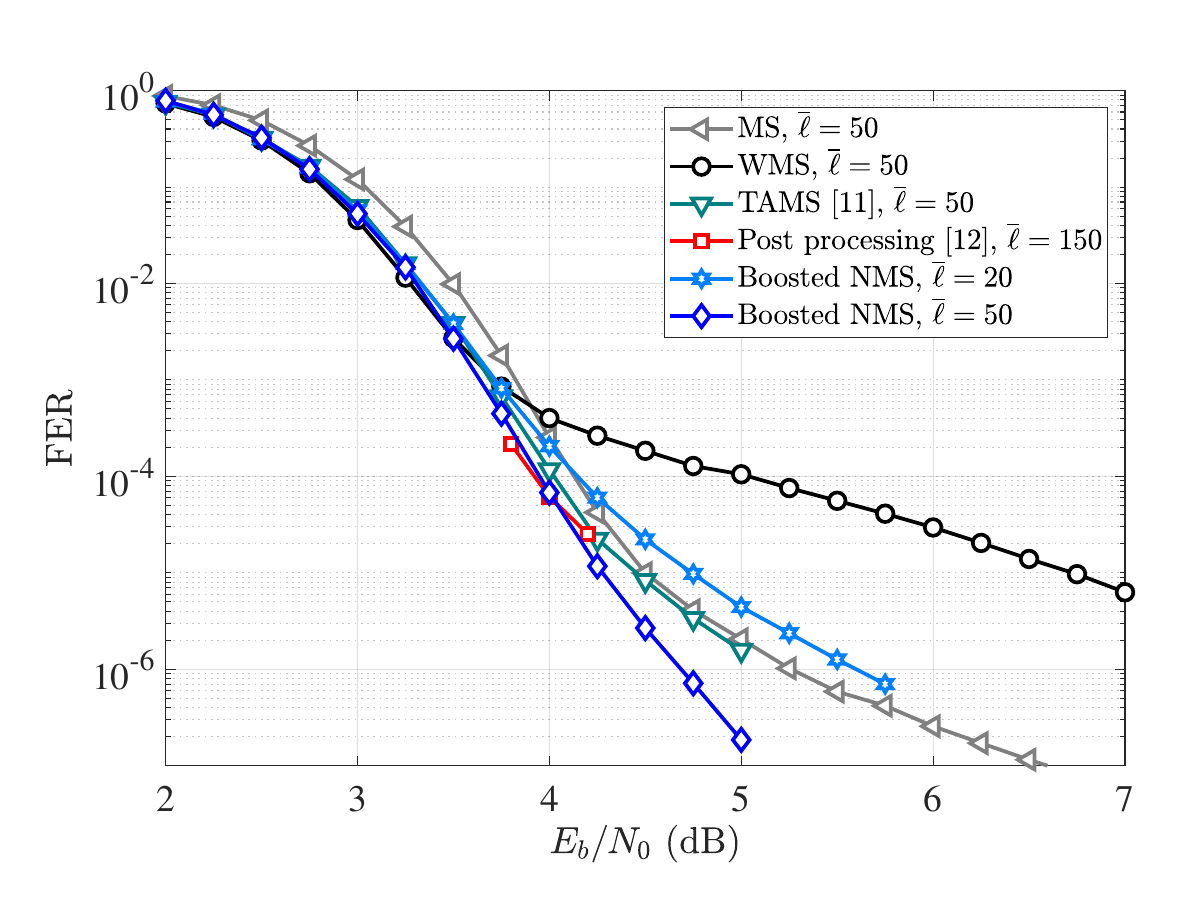}}
\hspace{10pt}
\subfigure[]{\includegraphics[scale=0.4]{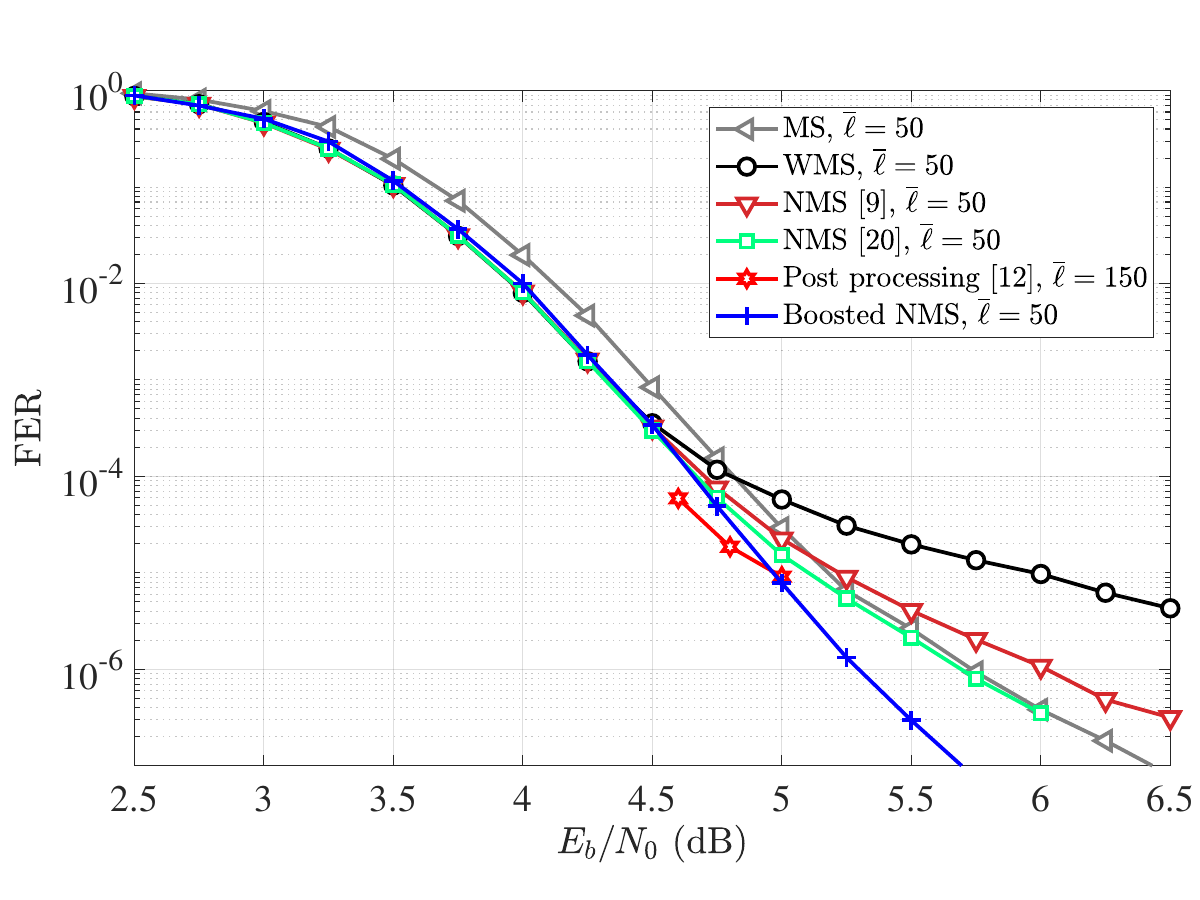}}
\caption{FER performances of (a): ($R=3/4, n=576$) WiMAX LDPC code, (b): ($R=5/6, n=648$) WiFi LDPC code.}
\label{Fig:FER_WiMax_WiFi}
\end{figure*}

Fig.~\ref{Fig:FER_WiMax_WiFi}(a) shows the FER performance of the boosted NMS decoder for the WiMAX LDPC code with $R=3/4, n=576$. The boosted NMS decoder is trained by the boosting learning, block-wise training, and dynamic weight sharing methods. The transfer learning and data augmentation methods are not utilized here, as the sampling process for the desired FER range up to $10^{-7}$ is manageable without the fast sampling methods. 

We compare the performance with the MS decoding, WMS decoding, offset MS (OMS) decoding, and other algorithm-based decoding approaches targeting the error floor \cite{Hatami2020, han2022deep}. The TAMS decoder \cite{Hatami2020} with the parameter set $\alpha=0.8,\tau=3.0$ shows a similar performance with the WMS and MS decoders in the waterfall and error floor regions, respectively.
We also compare with the state-of-the-art post processing scheme \cite{han2022deep}. 
Simulation results are directly referenced. While the post processing scheme \cite{han2022deep} shows comparable or slightly inferior performance to the boosted NMS decoder, it suffers from high decoding complexity and latency due to its large number of iterations $\overline{\ell}=150$. 
On the contrary, the boosted NMS decoder exhibits a barely noticeable error floor down to FER of $10^{-7}$ with $\overline{\ell}=50$ iterations.
In the comparison between $\overline{\ell}=20$ and $\overline{\ell}=50$ settings for the boosted NMS decoder, it is evident that the post decoder successfully mitigates the error floor.

Fig.~\ref{Fig:FER_WiMax_WiFi}(b) shows the FER performances for the WiFi LDPC code with $R=5/6, n=648$.
In this experiment, the boosted NMS decoder is compared with conventional NMS decoders \cite{nachmani2018deep, dai2021learning}. We exclude neural decoders that rely on different decoding algorithms or require the enumeration of trapping sets.
For the NMS decoders in \cite{nachmani2018deep, dai2021learning}, the base decoding stage is the same as that of the boosted NMS decoder, while the post decoding stage is trained by their respective training methods. 
The training method in \cite{nachmani2018deep} involves the use of received vectors from the waterfall region (${\rm E}_{\rm b}/{\rm N}_0$ $2$-$4$ dB) as training samples, training all weights all at once without a training schedule. 
For the training method in \cite{dai2021learning}, received vectors from the ${\rm E}_{\rm b}/{\rm N}_0$ points where the MS decoder achieves BER of $10^{-3}$ are used as training samples, and the iter-by-iter training schedule is employed. 
Both methods use the full weight set to achieve their best performance. 
The other hyper-parameters are aligned with those in our proposed method.
As shown in Fig. \ref{Fig:FER_WiMax_WiFi}(b), the conventional NMS decoders \cite{nachmani2018deep, dai2021learning} perform well in the waterfall region (${\rm E}_{\rm b}/{\rm N}_0$ $2$-$4$ dB), but start to exhibit the error floor beyond $4$ dB. In contrast, the boosted NMS decoder shows excellent performance in both the waterfall and error floor regions.

\begin{table*}[t]
\centering
\caption{Complexity comparison for the WiMAX LDPC code of $(N,M,E,z,\alpha)=(24,6,88,24,15.6)$, where $\alpha$ is the comparison count for each CN (i.e., $\alpha=d_c+\lceil {\rm log} d_c \rceil-2$ for the CN degree $d_c$ \cite{ryan2009channel})}
\begin{tabular}{c|ccc|c|c}
\hline
\multirow{2}{*}{}                                                                               & \multicolumn{3}{c|}{Complexity per iteration}                                                                                                                                                                                                                                      & Total complexity          & Memory      \\ \cline{2-4}
                                                                                                & \multicolumn{1}{c|}{Addition $A$}                                                             & \multicolumn{1}{c|}{Comparison $C$}                                                                 & Multiplication $M$                                                           & $(A+2C+M)\overline{\ell}$ & for weights \\ \hline \hline
MS, $\overline{\ell}=50$                                                                        & \multicolumn{1}{c|}{\multirow{6}{*}{\begin{tabular}[c]{@{}c@{}}$2Ez$\\ $=4224$\end{tabular}}} & \multicolumn{1}{c|}{\multirow{6}{*}{\begin{tabular}[c]{@{}c@{}}$\alpha Mz$\\ $=2256$\end{tabular}}} & $Ez=2112$                                                                    & $542400$                  & 0           \\
\cline{1-1} \cline{4-6}
WMS $\overline{\ell}=50$                                                                        & \multicolumn{1}{c|}{} & \multicolumn{1}{c|}{} & \multirow{2}{*}{\begin{tabular}[c]{@{}c@{}}$2Ez=4224$\end{tabular}}                                                                    & \multirow{2}{*}{\begin{tabular}[c]{@{}c@{}}$648000$\end{tabular}}                  & 1           \\
\cline{1-1} \cline{6-6}
TAMS \cite{Hatami2020} $\overline{\ell}=50$                                                                        & \multicolumn{1}{c|}{} & \multicolumn{1}{c|}{} &                                                                     &                   & 2           \\

\cline{1-1} \cline{4-6} 
NMS \cite{nachmani2018deep, dai2021learning}, $\overline{\ell}=50$                                        & \multicolumn{1}{c|}{}                                                                         & \multicolumn{1}{c|}{}                                                                               & \multirow{2}{*}{\begin{tabular}[c]{@{}c@{}}$(2E+N)z$\\ $=4800$\end{tabular}} & \multirow{2}{*}{$676800$} & $3760$      \\ \cline{1-1} \cline{6-6} 
Boosted NMS, $\overline{\ell}=50$                                                              & \multicolumn{1}{c|}{}                                                                         & \multicolumn{1}{c|}{}                                                                               &                                                                              &                           & $130$       \\ \cline{1-1} \cline{4-6} 
\begin{tabular}[c]{@{}c@{}}Post processing \cite{han2022deep}\\ $\overline{\ell}=150$\end{tabular} & \multicolumn{1}{c|}{}                                                                         & \multicolumn{1}{c|}{}                                                                               & $Ez=2112$                                                                    & $1627200$                 & $0$         \\ \hline
\end{tabular}
\label{Table:Complexity}
\end{table*}

Table \ref{Table:Complexity} offers a comparison of the decoding complexity for the WiMAX LDPC code. 
The boosted NMS decoder has additional multiplications for the weighting operation, requiring $(E+N)z$ more than the MS decoder and $Nz$ more than the WMS and TAMS \cite{Hatami2020} decoders. 
The number of other operations remains unchanged. The overall complexity is evaluated with the assumption that the comparison operation $C$ is twice as complex as the addition $A$ and multiplication $M$ \cite{gamage2017channel}. The boosted NMS decoder requires additional memory for storing $2\ell_1+3\ell_2$ weights, whose number is less than the number of weights for the conventional NMS decoders \cite{nachmani2018deep, dai2021learning}. 
Since the post-processing scheme in \cite{han2022deep} does not use weighting, the complexity per iteration is lower than that of the boosted NMS scheme. However, the total complexity exceeds that of the proposed method by more than double due to the higher required number of iterations. Furthermore, additional complexity is introduced by the error path detector \cite{han2022deep}.

\subsection{Results for 5G LDPC Codes}
 \begin{figure*}[t]
\centering
\subfigure[]{\includegraphics[scale=0.4]{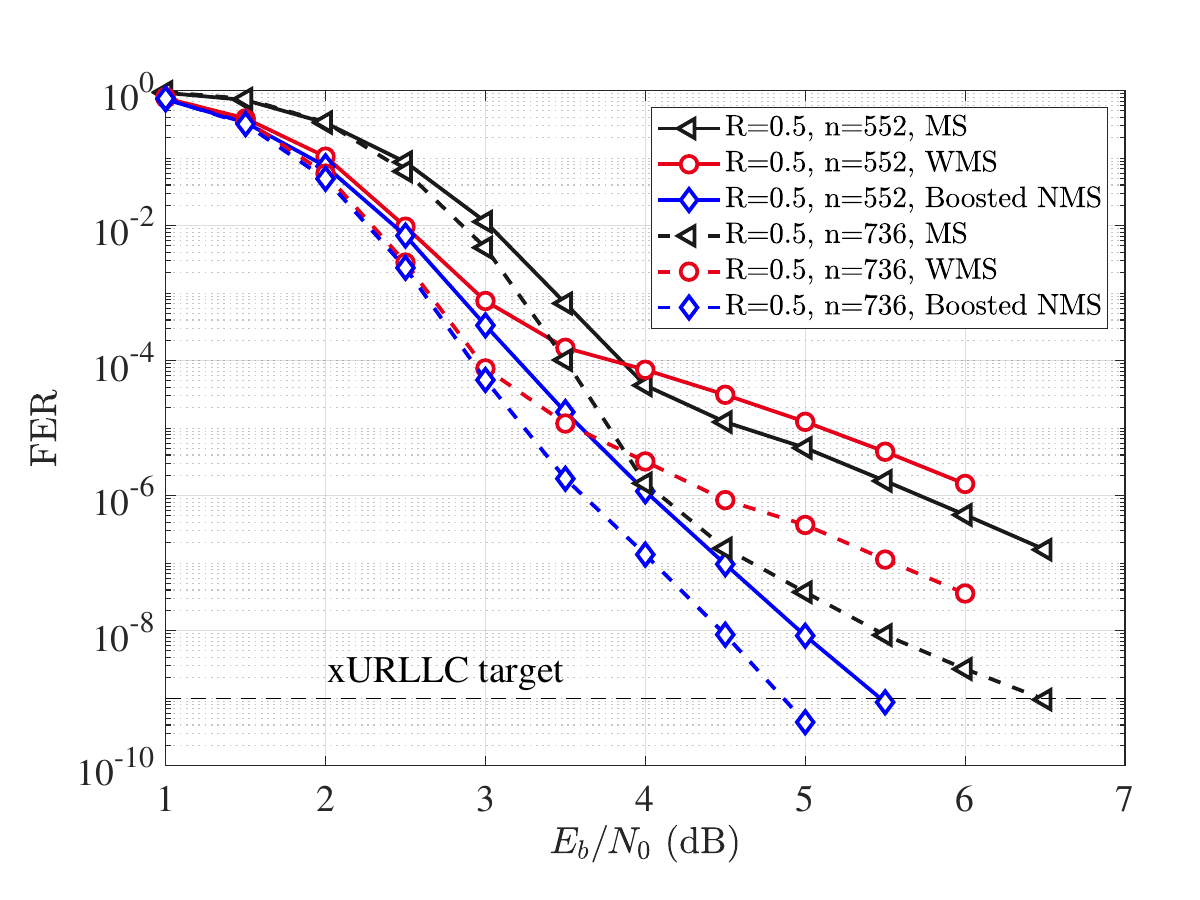}}
\hspace{10pt}
\subfigure[]{\includegraphics[scale=0.4]{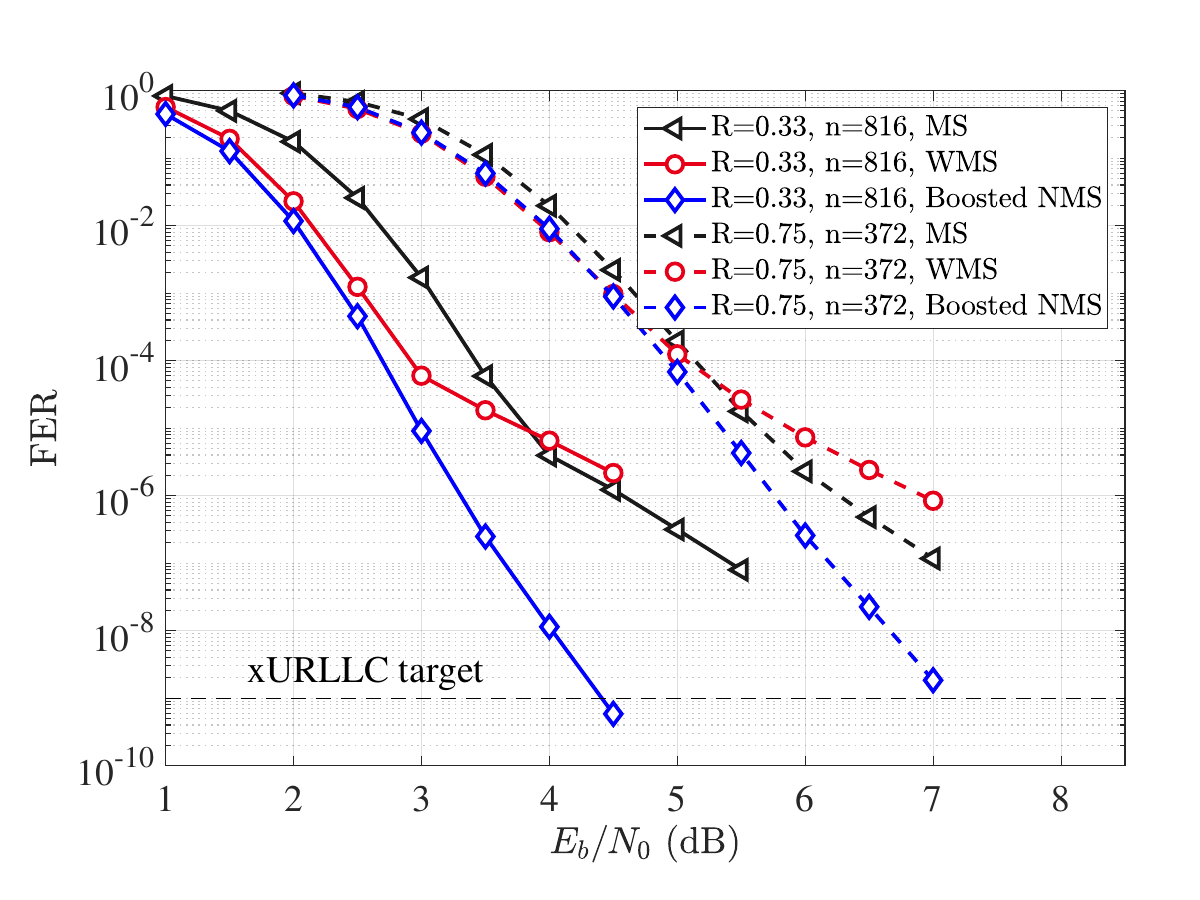}}
\subfigure[]{\includegraphics[scale=0.4]{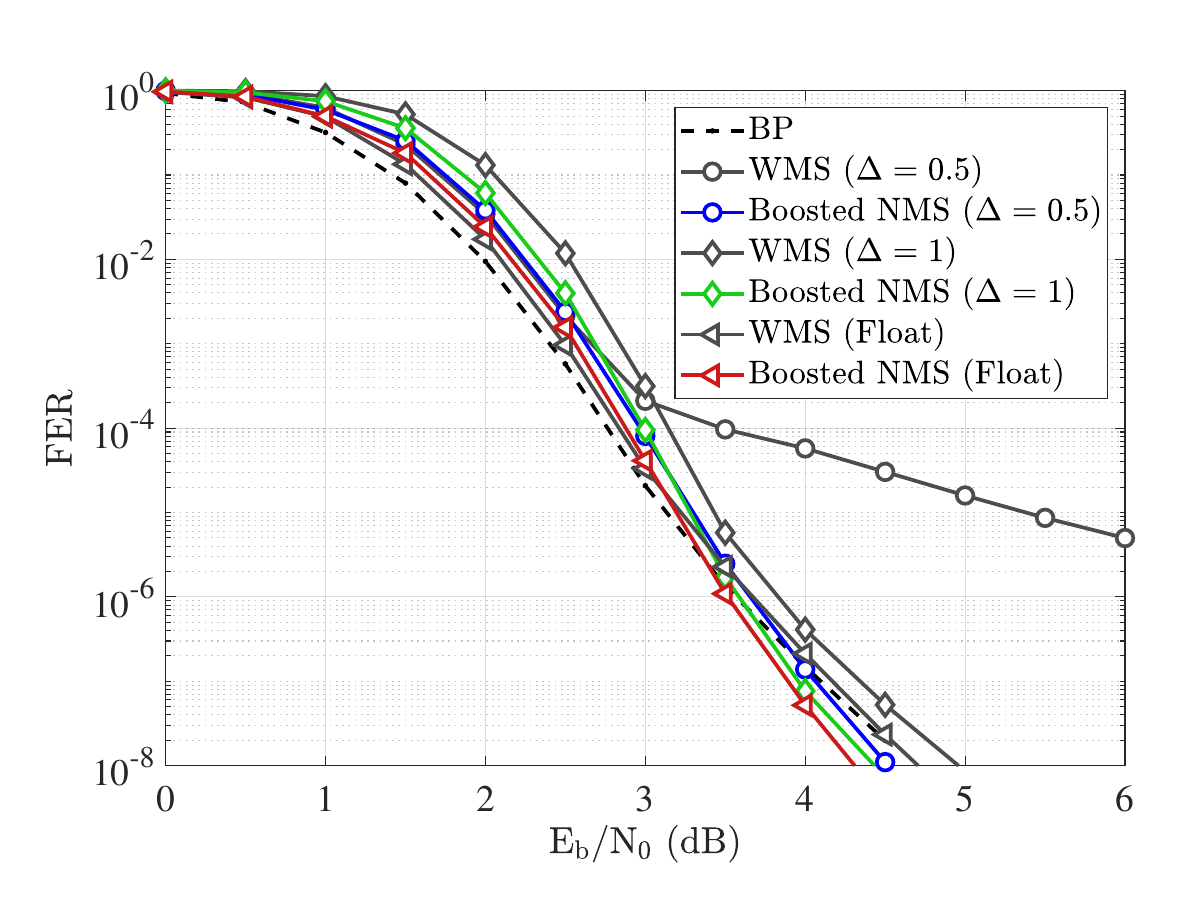}}
\hspace{10pt}
\subfigure[]{\includegraphics[scale=0.4]{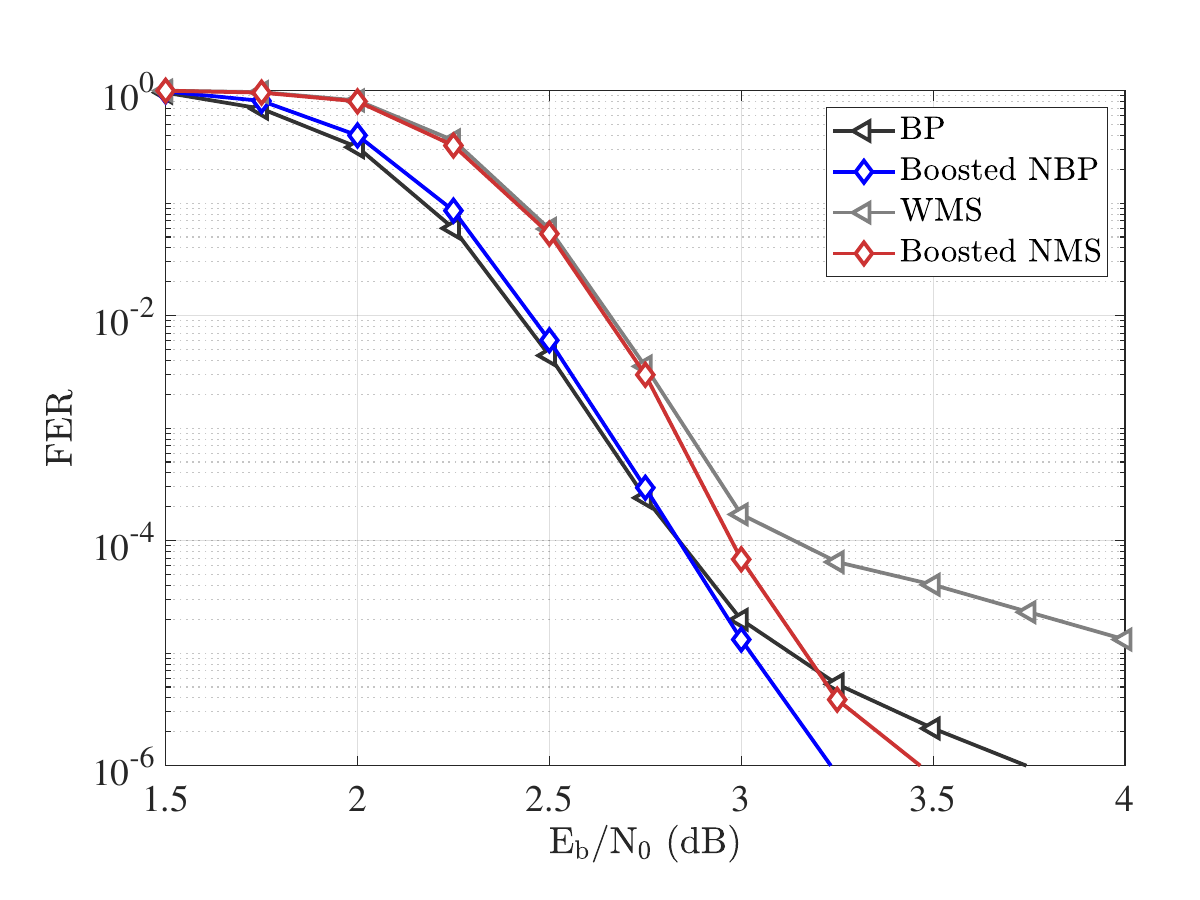}}
\caption{FER performances of (a): $(512,256)$ and $(1024,512)$ 5G LDPC codes, (b): $(786,256)$ and $(352,256)$ 5G LDPC codes, (c): $(512,256)$ 5G LDPC code with various message representations and (c): $(2112,1536)$ 5G LDPC code.}
\label{Fig:FER_5G_LDPC}
\end{figure*}

In Figs.\mbox{~\ref{Fig:FER_5G_LDPC}}(a) and (b), we evaluate the performance of four $(n,k)$ 5G LDPC codes with the following specifications: $(512,256)$, $(1024,512)$, $(786,256)$, $(352,256)$, whose code rates are $0.5,0.5,0.33,0.73$, respectively. For 5G LDPC codes that exhibit weak protection in the parity parts, we specifically evaluate FER for the information parts only.
To train the boosted NMS decoder that achieves the xURLLC FER target of $10^{-9}$, we utilize the methods of transfer learning and data augmentation. 
The total $50$ iterations are divided into $20$ for the base stage, and $15$ each for the first and second post stages. Since the first post decoder with only $15$ iterations is able to correct most UC vectors from the base decoder, we do not allocate additional iterations nor employ the block-wise training schedule. Instead, the second post decoder is used to correct UC vectors from the base and first post decoders.
In training the second post decoder, we use a more flexible weight set $\{\overline{w}^{(\ell)}_{v_p}, w^{(\ell)}_{c_p},{\hat {w}}^{(\ell)}_{c_p}\}_{\ell}$ to further improve performance.

The base decoder is trained in the waterfall region as usual. For the $(512,256)$ code, the first post decoder is trained using $10{,}000$ UC vectors without applying transfer learning. 
We denote the trained weights as baseline weights. 
These baseline weights are then used in transfer learning for training the second post decoder for the $(512,256)$ code, as well as the first and second post decoders for the other codes. For data augmentation, $100$ UC vectors are collected at the onset of the error floor region and augmented into $1{,}000$ UC vectors.

Figs.\mbox{~\ref{Fig:FER_5G_LDPC}}(a) and (b) demonstrate that the boosted NMS decoder achieves the xURLLC target across all evaluated codes without encountering any severe error floor. For example, with the $(512,256)$ code, the boosted NMS decoder reaches the target FER $10^{-9}$ at $5$ dB, while the WMS decoder only achieves FER of $10^{-5}$ at the same $5$ dB, implying $10^{4}$ gain by the boosted NMS decoder.
In addition, it shows superior performance in the waterfall region as well.
For other codes, the boosted NMS decoder also achieves the target FER at lower ${\rm E}_{\rm b}/{\rm N}_0$ values. Owing to the accelerated sampling speed by transfer learning and data augmentation, it is possible to train the boosted NMS decoder to achieve the extremely low error rate of $10^{-9}$ across various codes.

One reason the WMS decoder exhibits a severe error floor starting from FER of $10^{-4}$, as seen in Figs.\mbox{~\ref{Fig:FER_5G_LDPC}}(a) and (b), is the quantization scheme. To investigate the impact of quantization, Fig.\mbox{~\ref{Fig:FER_5G_LDPC}}(c) presents the performance of different message representations. Up until now, we have assumed a quantization scheme with a step size of $\Delta=0.5$ to achieve finer message representation, which is effective in the waterfall region. However, this representation imposes a maximum LLR value of $7.5$, leading to poor error floor performance. Increasing the step size to $1.0$ allows the maximum LLR value to rise to $15.0$, resulting in a more gradual error floor starting from $10^{-5}$, though at the cost of degraded waterfall performance, as shown in Fig.\mbox{~\ref{Fig:FER_5G_LDPC}}(c). If quantization is eliminated and floating-point message representations are used instead, both the waterfall and error floor performances improve significantly, but this is impractical for hardware implementation.
In contrast to this trade-off for the WMS decoder, the boosted NMS decoder shows a balanced performance in both the waterfall and error floor regions for all message representations. Notably, the quantized boosted NMS decoder (blue and green curves) nearly aligns with the boosted NMS decoder using floating-point implementation (red curve) and outperforms the BP decoder at low error rates, implying that the negative effects of quantization are mitigated by boosting learning. This is of significant practical importance, as quantization is essential for cost-effective hardware implementation.

Fig.\mbox{~\ref{Fig:FER_5G_LDPC}}(d) shows the performances of the $(2112,1536)$ long 5G LDPC code, where the BP decoder operating with floating-point message representation exhibits an error floor below $10^{-5}$. Although 5G LDPC codes are designed to avoid an error floor above FER of $10^{-4}$ with the BP decoder \mbox{\cite{Richardson2018}}, they do not guarantee the absence of an error floor below this point. When the proposed techniques are applied to the BP algorithm, resulting in the boosted NBP decoder, the error floor is almost entirely mitigated. Similarly, the boosted NMS decoder also achieves improved error floor performance compared to the WMS decoder. This result demonstrates that the proposed methods can be broadly applied to various decoding algorithms.

\subsection{Comparison with Augmented Neural Decoders}
\label{subsec:Short}

\begin{table*}[t]
\centering
\caption{Comparison with \cite{nachmani2018deep,choukroun2022error,nachmani2019hyper} in terms of waterfall performance of short codes (The results are
measured by the negative natural logarithm of BER. Higher is better.)}
\setlength{\tabcolsep}{4pt}
\begin{tabular}{c|ccccccccc|ccc|cccccc}
\hline
Architecture                & \multicolumn{9}{c|}{Standard NBP}                                                                                                 & \multicolumn{3}{c|}{NBP+HyperNet.}                  & \multicolumn{6}{c}{Transformer}                                                                          \\ \hline\hline
\multirow{2}{*}{Method}     & \multicolumn{3}{c|}{NBP \cite{nachmani2018deep}} & \multicolumn{3}{c|}{Boosted NBP}          & \multicolumn{3}{c|}{Boosted NBP}  & \multicolumn{3}{c|}{Hyper \cite{nachmani2019hyper}} & \multicolumn{3}{c|}{ECCT \cite{choukroun2022error}} & \multicolumn{3}{c}{ECCT \cite{choukroun2022error}} \\
                            & \multicolumn{3}{c|}{($\overline{\ell}=50$)}                 & \multicolumn{3}{c|}{($\overline{\ell}=50$)}          & \multicolumn{3}{c|}{($\overline{\ell}=100$)} & \multicolumn{3}{c|}{($\overline{\ell}=50$)}                    & \multicolumn{3}{c|}{($N=6$)}                        & \multicolumn{3}{c}{($N=10$)}                       \\ \hline
${\rm E}_{\rm b}/{\rm N}_0$ & 4       & 5       & \multicolumn{1}{c|}{6}       & 4    & 5     & \multicolumn{1}{c|}{6}     & 4         & 5         & 6         & 4               & 5                & 6              & 4        & 5        & \multicolumn{1}{c|}{6}        & 4              & 5               & 6               \\ \hline
Polar (64,48)               & 4.70    & 5.93    & \multicolumn{1}{c|}{7.55}    & 4.93 & 6.64  & \multicolumn{1}{c|}{8.77}  & -         & -         & -         & 4.92            & 6.44             & 8.39           & 6.36     & 8.46     & \multicolumn{1}{c|}{11.09}    & -              & -               & -               \\
BCH (63,51)                 & 4.64    & 6.21    & \multicolumn{1}{c|}{8.21}    & 4.72 & 6.42  & \multicolumn{1}{c|}{8.96}  & -         & -         & -         & 4.80            & 6.44             & 8.58           & 5.66     & 7.89     & \multicolumn{1}{c|}{11.01}    & -              & -               & -               \\
MacKay (96,48)              & 8.66    & 11.52   & \multicolumn{1}{c|}{14.32}   & 8.26 & 11.83 & \multicolumn{1}{c|}{15.85} & 8.52      & 12.48     & 16.88     & 8.90            & 11.97            & 14.94          & 7.23     & 10.42    & \multicolumn{1}{c|}{14.12}    & 8.39           & 12.24           & 16.41           \\ \hline
\end{tabular}
\label{Table:Waterfall}
\end{table*}

In Table~\ref{Table:Waterfall}, we compare with augmented neural decoders \cite{choukroun2022error,nachmani2019hyper} for short length codes including polar, BCH, MacKay LDPC codes in terms of waterfall performance. To ensure a fair comparison with other works, we employ the boosted NBP decoder and utilize the soft-BER loss function for BER performance optimization while making use of a general weight set $\{\overline{w}_{v}^{(\ell)}, w_{c \rightarrow v}^{(\ell)}, \hat{w}_{c \rightarrow v}^{(\ell)}\}_{\ell}$. 

Table~\ref{Table:Waterfall} shows a notable performance enhancement over the original NBP decoder \cite{nachmani2018deep}, which employs the standard NBP architecture. The improvement becomes more significant at higher SNR levels, where the boosted NBP decoder even outperforms the decoder \cite{nachmani2019hyper} that incorporates hyper-networks into the model architecture. When compared to the ECCT \cite{choukroun2022error} using the transformer architecture, our method shows inferior performance for high density codes (Polar, BCH codes), but achieves fairly comparable performance for LDPC codes. Notably, the boosted NBP decoder with $\overline{\ell}=100$ outperforms the ECCT with $N=10$ for the MacKay LDPC code, where $N$ refers to the number of layers in the transformer architecture. 

\begin{figure}[t]
\centering
\includegraphics[scale=0.45]{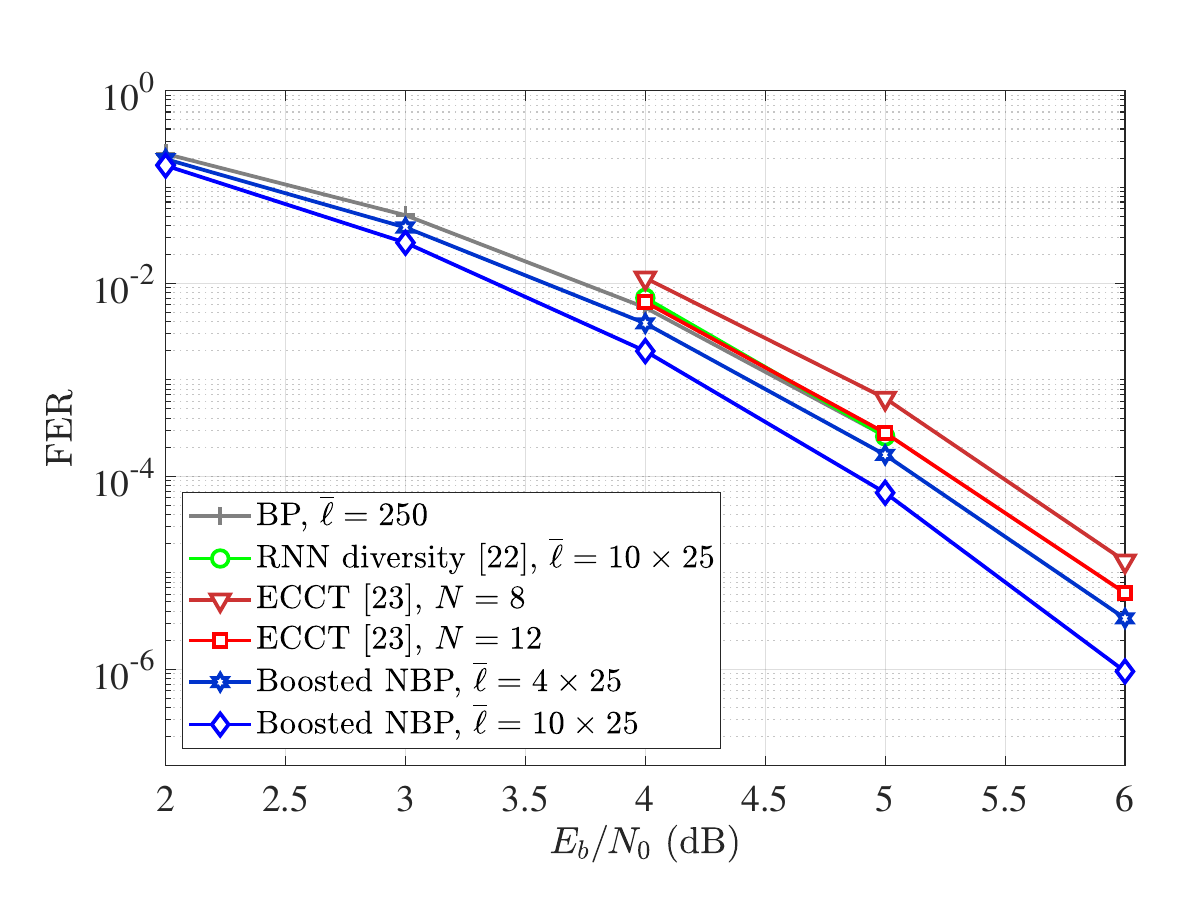}
\caption{FER performances of a short LDPC code of length $128$.}
\label{Fig:FER_Short_LDPC}
\end{figure}

In Fig.~\ref{Fig:FER_Short_LDPC}, the boosted NBP decoder is compared with the ECCT \cite{choukroun2022error} as well as the BP-RNN decoder \cite{rosseel2022decoding} for a short LDPC code of length $128$, which is provided in \cite{rosseel2022decoding}. The BP-RNN decoder employs noisy codeword vectors generated with the knowledge of absorbing sets for its training samples. The performance data is taken directly from \cite{rosseel2022decoding}. For a fair comparison, the boosted NMS decoder is configured similarly to the parallel BP-RNN decoder \cite{rosseel2022decoding} where the total of $\overline{\ell}=250=10\times 25$ iterations is segmented into $10$ parallel stages, each with $25$ iterations. Thus, it can be performed with low latency decoding in parallel. This configuration includes $1$ base decoder and $9$ post decoders in the context of boosting learning. The result shows that the boosted NBP decoder with $\overline{\ell}=10\times25$ surpasses both the BP-RNN decoder \cite{rosseel2022decoding} and ECCT \cite{choukroun2022error}. This underlines the effectiveness of the proposed training methodology not just in the error floor region of long LDPC codes, but also in the waterfall region of short LDPC codes.

Even though the comparison in this subsection considers augmented neural networks as competitors, we want to note that our proposed training methodology is adaptable regardless of decoding architectures. This flexibility opens up promising avenues for future research, particularly in terms of integrating our training techniques with advanced architectures such as hyper-networks and the transformer architectures. Such integration could potentially enhance the decoder performance for various code types.

\section{Conclusions}\label{Sec:Conclusion}
In this paper, we propose a set of training methodologies for the boosted NMS decoder, aimed at tackling the error floor$-$a longstanding problem of LDPC codes. The core of our methods lies in boosting learning, which utilizes selectively filtered received vectors (i.e., UC vectors) as training samples. These samples are particularly effective at mitigating the error floor, yet they seldom appear in conventional learning methods. Moreover, we propose supplementary methods to enhance the proposed boosting learning method, including: the block-wise training technique that enables efficient training over long iterations, the dynamic weight sharing technique for achieving superior performance with fewer weights, and transfer learning combined with data augmentation to expedite the sampling process of boosting learning. Through these complementary training methods, we developed the boosted NMS decoder capable of significantly lowering the error floor and achieving performance levels suitable for 6G networks. The proposed training techniques have broad applicability across different types of channels, codes, and decoding algorithms, providing substantial implications for any domain demanding high reliability. The boosted NMS decoder is also practical and readily deployable in real-world applications because it achieves the performance improvement solely through weight diversification.




\begin{thebibliography}{11}

\bibitem{Tataria2021}
H. Tataria, M. Shafi, A. F. Molisch, M. Dohler, H. Sjöland, and F. Tufvesson, ``6G Wireless Systems: Vision, requirements, challenges, insights, and opportunities," \emph{Proceedings of the IEEE}, vol. 109, no. 7, pp. 1166--1199, Jul. 2021.

\bibitem{hong2022}
E.-K. Hong, I. Lee, B. Shim, Y.-C. Ko, S.-H. Kim, S. Pack, K. Lee, S. Kim, J.-H. Kim, Y. Shin, Y. Kim, and H. Jung, ``6G R\&D vision: Requirements and candidate technologies,'' \emph{J. Commun. Netw.}, vol. 24, no. 2, pp. 232--245, 2022.

\bibitem{Yang2021}
X. Yang, Z. Zhou, and B. Huang, ``URLLC key technologies and standardization for 6G power internet of things," \emph{IEEE Commun. Stand. Mag.}, vol. 5, pp. 52--59, 2021.


\bibitem{dong2010soft}
G. Dong, N. Xie, and T. Zhang, ``On the use of soft-decision error-correction codes in NAND flash memory,'' \emph{IEEE Trans. Circuits Syst. I: Reg. Papers}, vol. 58, no. 2, pp. 429--439, 2010.

\bibitem{chandak2019improved}
S. Chandak, K. Tatwawadi, B. Lau, J. Mardia, M. Kubit, J. Neu, P. Griffin, M. Wootters, T. Weissman, and H. Ji, ``Improved read/write cost tradeoff in DNA-based data storage using LDPC codes,'' in \emph{Proc. Allerton Conf. Commun. Control Comput. (Allerton)}, 2019, pp. 147--156

\bibitem{baldi2008new}
M. Baldi, M. Bodrato, and F. Chiaraluce, ``A new analysis of the McEliece cryptosystem based on QC-LDPC codes,'' in \emph{Proc. Security and Cryptography for Networks (SCN)}, Springer, Sep., 2008, pp. 246--262


\bibitem{Richardson2003}
T. J. Richardson, “Error floors of LDPC codes,” in {\em Proc. Allerton Conf. Commun. Control Comput. (Allerton)}, Oct. 2003, pp. 1426--1435.


\bibitem{zhang2014quantized}
X. Zhang and P. H. Siegel, ``Quantized iterative message passing decoders with low error floor for LDPC codes,'' \emph{IEEE Trans. Commun.}, vol. 62, no. 1, pp. 1--14, 2014.

\bibitem{nachmani2018deep}
E. Nachmani, E. Marciano, L. Lugosch, W. J. Gross, D. Burshtein, and Y. Be’ery, ``Deep learning methods for improved decoding of linear codes,'' \emph{IEEE J. Sel. Topics Signal Process.}, vol. 12, no. 1, pp. 119--131, 2018.



\bibitem{freund1997decision}
Y. Freund and R. E. Schapire, ``A decision-theoretic generalization of on-line learning and an application to boosting,'' \emph{J. Comput. Syst. Sci.}, vol. 55, no. 1, pp. 119--139, 1997.


\bibitem{Hatami2020}
H. Hatami, D. G. M. Mitchell, D. J. Costello, and T. E. Fuja, ``A threshold-based min-sum algorithm to lower the error floors of quantized LDPC decoders,'' {\em IEEE Trans. Commun.}, vol. 68, no. 4, pp. 2005--2015, Apr. 2020.

\bibitem{han2022deep}
S. Han, J. Oh, K. Oh, and J. Ha, ``Deep-learning for breaking the trapping sets in low-density parity-check codes,'' \emph{IEEE Trans. Commun.}, vol. 70, no. 5, pp. 2909--2923, May, 2022.

\bibitem{lee2022post} H. Lee, Y.-S. Kil, M. Y. Chung, and S.-H. Kim, ``Neural network aided impulsive perturbation decoding for short raptor-like LDPC codes,'' \emph{IEEE Wireless Commun. Lett.}, vol. 11, no. 2, pp. 268--272, Feb. 2022. 

\bibitem{glorot2010understanding}
X. Glorot and Y. Bengio, ``Understanding the difficulty of training deep feedforward neural networks,'' in \emph{Proc. Int. Conf. Artif. Intell. Stat.}, 2010, pp. 249--256.

\bibitem{Dolecek2009}
L. Dolecek, P. Lee, Z. Zhang, V. Anantharam, B. Nikolic, and
M. Wainwright, “Predicting error floors of structured LDPC codes:
Deterministic bounds and estimates,” \emph{IEEE J. Sel. Areas Commun.},
vol. 27, no. 6, pp. 908--917, Jun. 2009.


\bibitem{ieee80216e2020}
IEEE, ``IEEE standard for local and metropolitan area networks-part 16: Air interface for fixed and mobile broadband wireless access systems amendment 2: Physical and medium access control layers for combined fixed and mobile operation in licensed bands and corrigendum 1,'' \emph{IEEEStandard 802.16e}, 2020.

\bibitem{ieee802112016}
IEEE, ``IEEE standard for information technology—telecommunications and information exchange between systems local and metropolitan area networks—specific requirements—part 11: Wireless lan medium access control (MAC) and physical layer (PHY) specification,'' \emph{IEEE Standard 802.11}, 2016.

\bibitem{3gpp2023}
3rd Generation Partnership Project, ``Technical specification group radio access network; NR; multiplexing and channel coding (release 17) document ts 38.212 v17.5.0,'' 2023.

\bibitem{chen2005reduced}
J. Chen, A. Dholakia, E. Eleftheriou, M. P. C. Fossorier, and X.-Y. Hu, ``Reduced-complexity decoding of LDPC codes,'' \emph{IEEE Trans. Commun.}, vol. 53, no. 8, pp. 1288--1299, 2005.

\bibitem{dai2021learning}
J. Dai, K. Tan, Z. Si, K. Niu, M. Chen, H. V. Poor, and S. Cui, ``Learning to decode protograph LDPC codes,'' \emph{IEEE J. Sel. Areas Commun.}, vol. 39, no. 7, pp. 1983--1999, 2021.

\bibitem{Kwak2023}
H. Kwak, D.-Y. Yun, Y. Kim, S.-H. Kim, and J.-S. No, "Boosting learning for LDPC codes to improve the error-floor performance," in {\em Proc. Adv. Neural Inf. Process. Syst. (NeurIPS)}, Dec. 2023. 


\bibitem{rosseel2022decoding}
J. Rosseel, V. Mannoni, I. Fijalkow, and V. Savin, ``Decoding short LDPC codes via BP-RNN diversity and reliability-based post-processing,'' \emph{IEEE Trans. Commun.}, vol. 70, no. 12, pp. 7830--7842, 2022.



\bibitem{choukroun2022error}
Y. Choukroun and L. Wolf, ``Error correction code transformer,'' in \emph{Proc. Adv. Neural Inf. Process. Syst. (NeurIPS)}, vol. 35, 2022, pp. 38695–38705.

\bibitem{Richter2006}
G. Richter and A. Hof, “On a construction method of irregular LDPC
codes without small stopping sets,” in {\em Proc. IEEE Int. Conf. Commun.},
vol. 3, Jun. 2006, pp. 1119–1124.

\bibitem{Jiao2009}
X. Jiao, J. Mu, J. Song, and L. Zhou, “Eliminating small stopping sets in
irregular low-density parity-check codes,” {\em IEEE Commun. Lett.}, vol. 13,
no. 6, pp. 435–437, Jun. 2009.

\bibitem{Asvadi2011}
R. Asvadi, A. H. Banihashemi, and M. Ahmadian-Attari, “Lowering
the error floor of LDPC codes using cyclic liftings,” {\em IEEE Trans. Inf. Theory}, vol. 57, no. 4, pp. 2213--2224, Apr. 2011.

\bibitem{Naseri2019}
S. Naseri and A. H. Banihashemi, ``Construction of Girth-8 QC-LDPC
codes free of small trapping sets,'' {\em IEEE Commun. Lett.}, vol. 23, no. 11, pp. 1904–1908, Nov. 2019.

\bibitem{Karimi2021}
B. Karimi and A. H. Banihashemi, “Construction of irregular
protograph-based QC-LDPC codes with low error floor,” {\em IEEE Trans.
Commun}., vol. 69, no. 1, pp. 3–18, Jan. 2021.


\bibitem{caid1990neural}
W. R. Caid and R. W. Means, ``Neural network error correcting decoders for block and convolutional codes,'' in \emph{Proc. IEEE Global Telecommunications Conference and Exhibition (GLOBECOM)}, 1990, pp. 1028--1031.

\bibitem{tallini1995neural}
L. G. Tallini and P. Cull, ``Neural nets for decoding error-correcting codes,'' in \emph{Proc. IEEE Tech. Appl. Conf. Workshops}, 1995, p. 89.

\bibitem{gruber2017deep}
T. Gruber, S. Cammerer, J. Hoydis, and S. ten Brink, ``On deep learning-based channel decoding,'' in \emph{Proc. Annu. Conf. Inf. Sci. Syst. (CISS)}, 2017, pp. 1--6.

\bibitem{Lee2020} H. Lee, E. Y.  Seo, H. Ju, and S.-H. Kim ``On training neural network decoders of rate compatible polar codes via transfer learning'' \emph{Entropy}, vol. 22, no. 5, 2020.

\bibitem{Bennatan2018}
A. Bennatan, Y. Choukroun, and P. Kisilev, “Deep learning for decoding
of linear codes—A syndrome-based approach,” in \emph{Proc. IEEE Int. Symp.
Inf. Theory (ISIT)}, 2018, pp. 1595–1599.

\bibitem{Jiang2020}
Y.~Jiang, H.~Kim , H.~Asnani, S.~Kannan, S.~Oh, and P.~Viswanath,
``LEARN codes: Inventing low-latency codes via recurrent neural networks,'' \emph{IEEE J. Sel. Areas Commun.}, vol. 1, no. 1, pp. 207--216, May 2020.

\bibitem{O’Shea2017}
T. O’Shea and J. Hoydis, ``An introduction to deep learning for
the physical layer,'' IEEE Trans. Cogn. Commun., vol. 3, no. 4,
pp. 563--575, Dec. 2017.

 \bibitem{Jamali2022}
M. V. Jamali, H. Saber, H. Hatami, and J. H. Bae, ``ProductAE:
Toward training larger channel codes based on neural product
codes,'' in \emph{Proc. IEEE Int. Conf. Commun.}, 2021, pp. 3898--3903

\bibitem{Clausius2023}
J. Clausius, M. Geiselhart, and S. ten Brink, “Component training of turbo autoencoders,” in \emph{Proc. International Symposium on Topics in Coding (ISTC)}, 2021.

\bibitem{lian2019learned}
M. Lian, F. Carpi, C. Häger, and H. D. Pfister, ``Learned belief-propagation decoding with simple scaling and SNR adaptation,'' in \emph{Proc. IEEE Int. Symp. Inf. Theory (ISIT)}, 2019, pp. 161--165.

\bibitem{Kumar2023}
S. K. Ankireddy and H. Kim, ``Interpreting neural min-sum decoders,'' in \emph{Proc. IEEE Conference on Communications (ICC)}, 2023, pp. 6645--6651.

\bibitem{Wang2024}
L. Wang, C. Terrill, D. Divsalar, and R. D. Wesel, ``LDPC decoding With degree-specific neural message weights and RCQ decoding,'' \emph{IEEE Trans. Commun.}, vol. 72, no. 4, Apr., pp. 1912--1924, 2024.

\bibitem{Andreev2021}
K. Andreev, A. Frolov, G. Svistunov, K. Wu, and J. Liang, ``Deep neural network based decoding of short 5G LDPC codes,'' in \emph{Proc. XVII International Symposium Problems of Redundancy in Information and Control Systems}, 2021, pp. 155–160.

\bibitem{Kwak2022}
H. Kwak, J.-W. Kim, Y. Kim, S.-H. Kim, and J.-S. No, "Neural min-sum decoding for generalized LDPC codes," {\em IEEE Commun. Lett.}, vol. 26, no. 12, pp. 2841--2845, 2022.

\bibitem{Chen2021}
X. Chen and M. Ye, "Cyclically equivariant neural decoders for cyclic codes," in {\em Proc. International Conference on Machine Learning (ICML)}, 2021.

\bibitem{xiao2020designing}
X. Xiao, B. Vasic, R. Tandon, and S. Lin, ``Designing finite alphabet iterative decoders of LDPC codes via recurrent quantized neural networks,'' \emph{IEEE Trans. Commun.}, vol. 68, no. 7, pp. 3963--3974, 2020.

\bibitem{xiao2021faid}
X. Xiao, N. Raveendran, B. Vasic, S. Lin, and R. Tandon, ``FAID diversity via neural networks,'' in \emph{Proc. Int. Symp. Topics Coding (ISTC)}, 2021, pp. 1--5.

\bibitem{shah2021neural}
N. Shah and Y. Vasavada, ``Neural layered decoding of 5G LDPC codes,'' \emph{IEEE Commun. Lett.}, vol. 25, no. 11, pp. 3590--3593, 2021.

\bibitem{Artemasov2023}
D. Artemasov, K. Andreev, P. Rybin, and A. Frolov, ``Soft-output deep neural network-based decoding,'' in \emph{Proc. IEEE Globecom Workshops}, 2023, pp. 1692–1697.

\bibitem{nachmani2019hyper}
E. Nachmani and L. Wolf, ``Hyper-graph-network decoders for block codes,'' in \emph{Proc. Adv. Neural Inf. Process. Syst. (NeurIPS)}, vol. 32, 2019.

\bibitem{Beery2020}
I. Be’ery, N. Raviv, T. Raviv, and Y. Be’ery, “Active deep decoding of linear codes,” \emph{IEEE Trans. Commun.}, vol. 68, no. 2, pp. 728--736, 2020.

\bibitem{Buchberger2021}
A. Buchberger, C. Hager, H. D. Pfister, L. Schmalen, and A. Graell I Amat, "Pruning and quantizing neural belief propagation decoders," \emph{IEEE J. Sel. Areas Commun.}, vol. 39, no. 7, pp. 1957--1966, 2020.

\bibitem{Nachmani2022}
E. Nachmani and Y. Be’ery, ``Neural decoding with optimization of
node activations,'' \emph{IEEE Commun. Lett.}, vol. 26, no. 11, pp. 2527--2531, 2022.

\bibitem{ryan2009channel}
W. Ryan and S. Lin, \emph{Channel Codes: Classical and Modern}, Cambridge University Press, 2009.

\bibitem{fossorier2004quasicyclic}
M. P. C. Fossorier, ``Quasi cyclic low-density parity-check codes from circulant permutation matrices,'' \emph{IEEE trans. Inf. Theory}, vol. 50, no. 8, pp. 1788--1793, 2004.

\bibitem{thorpe2003low}
J. Thorpe, ``Low-density parity-check (LDPC) codes constructed from protographs,'' \emph{IPN progress report}, vol. 42, no. 154, pp. 42--154, 2003.

\bibitem{wu2010adaptive}
X. Wu, Y. Song, M. Jiang, and C. Zhao, ``Adaptive-normalized/offset min-sum algorithm,'' \emph{IEEE Commun. Lett.}, vol. 14, no. 7, pp. 667--669, 2010.

\bibitem{kingma2014adam}
D. P. Kingma and J. Ba, ``Adam: A method for stochastic optimization,'' \emph{arXiv preprint arXiv:1412.6980}, 2014.

\bibitem{Matsumoto1998}
M. Matsumoto and T. Nishimura, "Mersenne Twister: A 623-dimensionally equidistributed uniform pseudo-random number generator," {\em ACM Trans. Model. Comput. Simul.}, vol. 8, no. 1, pp. 3--30, Jan. 1998.

\bibitem{gamage2017channel}
H. Gamage, N. Rajatheva, and M. Latva-Aho, ``Channel coding for enhanced mobile broadband communication in 5G systems,'' in \emph{Proc. Eur. Conf. Netw. Commun. (EuCNC)}, 2017, pp. 1--6.

\bibitem{Richardson2018}
T. Richardson and S. Kudekar, ``Design of Low-Density Parity Check Codes for 5G New Radio,'' \emph{IEEE Commun. Mag.}, vol. 56, no. 3, pp. 28--34, Mar 2018.






\end{thebibliography}
\end{document}